\documentclass[aps,pra,amssymb,twocolumn,amsmath,superscriptaddress,10pt,tightenlines]{revtex4-1}

\usepackage{amsmath,amssymb} 
\usepackage{mathtools} 
\usepackage{physics} 
\usepackage{bm} 
\usepackage{amsthm} 
\usepackage{ascmac} 

\usepackage{graphicx} 
\usepackage{here} 

\usepackage[breaklinks,  hyperfootnotes=true, pdfpagelabels, bookmarks, pageanchor]{hyperref}
\hypersetup{%
	colorlinks=true, linktocpage=true, pdfstartpage=1, pdfstartview=FitH, pdfborder={0 0 0},%
	breaklinks=true, pdfpagemode=UseNone, pageanchor=true, pdfpagemode=UseOutlines,%
	plainpages=false, bookmarksnumbered, bookmarksopen=true, bookmarksopenlevel=1,%
	hypertexnames=true, pdfhighlight=/O,
	urlcolor=blue, linkcolor=blue, citecolor=blue, 
}

\mathtoolsset{showonlyrefs,showmanualtags}

\usepackage{overpic}

\begin{document}

\title{Kibble--Zurek scaling in the quantum Ising chain with a time-periodic perturbation}

\author{Takayuki Suzuki}
\affiliation{National Institute of Information and Communications Technology, Nukui-Kitamachi 4-2-1, Koganei,
Tokyo 184-8795, Japan}
\author{Kaito Iwamura}
\affiliation{Department of Physics, Waseda University, Tokyo 169-8555, Japan}

\begin{abstract}
We consider the time-dependent transverse field Ising chain with time-periodic perturbations. Without perturbations, this model is one of the famous models that obeys the scaling in the adiabatic limit predicted by the quantum Kibble--Zurek mechanism (QKZM). However, it is known that when oscillations are added to the system, the non-perturbative contribution becomes larger and the scaling may break down even if the perturbation is small. Therefore, we analytically analyze the density of defects in the model and discuss how much the oscillations affect the scaling. As a result, although the non-perturbative contribution does not become zero in the adiabatic limit, the scaling does not change from the prediction of the QKZM. This indicates that the QKZM is robust to the perturbations.

\end{abstract}

\maketitle

\section{Introduction}\label{sec:intro}

The Kibble--Zurek mechanism (KZM) is a fundamental concept that explains the formation of topological defects during non-equilibrium phase transitions. The original theory was proposed in the context of cosmology, where the universe underwent a symmetry-breaking phase transition in the early stages of its evolution~\cite{kibble1976topology,kibble1980some}. Since then, the KZM has been adapted to condensed matter systems, especially in the study of quantum phase transitions~\cite{zurek1985cosmological,zurek1993cosmic,
zurek1996cosmological}.
The KZM has been experimentally validated in a variety of systems, such as the superfluid helium experiments~\cite{hendry1994generation} and the superconductor experiments~\cite{monaco2002zurek,monaco2003spontaneous,maniv2003observation}. 

The quantum Kibble--Zurek mechanism (QKZM), is an extension of the KZM that incorporates quantum effects. In the context of phase transitions, quantum corrections can lead to significant modifications in the physics near the critical point, giving rise to novel phenomena. The QKZM has been developed to investigate how these quantum corrections affect the predictions of the KZM and take into account the quantum fluctuations near the critical point. The QKZM has already been studied~\cite{dziarmaga2010dynamics,
sinha2019kibble,
sadhukhan2020sonic,
dutta2017probing,
polkovnikov2011colloquium,
rossini2021coherent,
cincio2007entropy,
saito2007kibble,
sengupta2008exact,
sen2008defect,
dziarmaga2008winding,
divakaran2009reverse,
honer2010collective,
zurek2013topological,
cucchietti2007dynamics,
mukherjee2007quenching,
nag2013quenching,
del2014universality,
dutta2015quantum,
cherng2006entropy,
mukherjee2008defect,
rams2019symmetry,
revathy2020adiabatic,
rossini2020dynamic,
hodsagi2020kibble,
bialonczyk2020locating,
nowak2021quantum,
kou2022interferometry,
zurek2005dynamics,
kells2014topological,
heyl2013dynamical,
dziarmaga2005dynamics,
coldea2010quantum,
kinross2014evolution,
king2022coherent,
sachdev1999quantum,
sun2022universal,
zeng2023universal,
yan2021nonadiabatic,
fubini2007robustness,
bermudez2009topology,
mukherjee2009effects} and observed in many experiments~\cite{ulm2013observation,
pyka2013topological,
monaco2006zurek,
sadler2006spontaneous,
chen2011quantum,
griffin2012scaling,
lamporesi2013spontaneous,
navon2015critical,
braun2015emergence,
chomaz2015emergence,
anquez2016quantum,
clark2016universal,
keesling2019quantum,
baumann2011exploring,
xu2014quantum,
meldgin2016probing,
clark2016universal,
chen2020experimentally,
gardas2018defects,
gong2016simulating,
cui2016experimental,
bando2020probing,
li2023probing,
zamora2020kibble}.
In the study of the QKZM, a theoretical approach based on the one-dimensional transverse field Ising model is sometimes used~\cite{
nag2013quenching,
del2014universality,
revathy2020adiabatic,
cherng2006entropy,
mukherjee2007quenching,
mukherjee2008defect,
kells2014topological,
rams2019symmetry,
rossini2020dynamic,
hodsagi2020kibble,
bialonczyk2020locating,
nowak2021quantum,
kou2022interferometry,
zurek2005dynamics,
heyl2013dynamical,
dziarmaga2005dynamics,
coldea2010quantum,
kinross2014evolution,
king2022coherent,
sachdev1999quantum,
sun2022universal,
zeng2023universal,
yan2021nonadiabatic,
fubini2007robustness,
bermudez2009topology,
dutta2015quantum,
gardas2018defects,
chen2020experimentally,
bando2020probing,
cui2016experimental,
gong2016simulating,
li2023probing,
zamora2020kibble}. The system begins with all spins aligned which corresponds to the ground state at infinite past. As the system's energy evolves linearly with time such as $v(t-t_c)$, a phase transition occurs, resulting in the emergence of defects. According to the QKZM, the density of defects is generally given by $n \propto v^{d\nu/(1+z\nu)}$, where $d$ is the dimension of the system, $z$ is the dynamic exponent, and $\nu$ is the correlation length exponent. In the one-dimensional transverse field Ising model, these exponents are given by $z=\nu=1$~\cite{sachdev1999quantum,dziarmaga2010dynamics}.

This scaling is an estimate of the computational time for quantum annealing, since it corresponds to the probability of successfully obtaining the ground state. Therefore, it is important to investigate what happens to the scaling when the linear ramp is deviated or perturbations are added. The robustness to these changes has been investigated in several previous studies. For example, when the spin-spin coupling is changed alternately, the density of defects includes a factor that decays exponentially and is subject to large corrections~\cite{yan2021nonadiabatic}. Furthermore, the numerical simulation shows that the density of defects increases due to the effect of white noise.~\cite{fubini2007robustness}.
It is also known to exhibit nontrivial behavior when oscillations are added as perturbations, but the effect of the perturbation on the scaling is not derived analytically~\cite{mukherjee2009effects}. 

This nontrivial behavior, caused by adding an oscillation term to a linear ramp, is also observed in other fields. 
The Franz--Keldysh effect, originally proposed in the 1950s~\cite{franz1958einfluss,keldysh1958behavior,tharmalingam1963optical,callaway1963optical}, is an important phenomenon observed in semiconductors when subjected to strong electric fields. 
This analysis method is also applied to the dynamically assisted Schwinger mechanism~\cite{schutzhold2008dynamically}, the extension of the Schwinger mechanism~\cite{euler1936consequences,weisskopf1936k,schwinger1951gauge} which explains the phenomenon in quantum electrodynamics where electron-positron pairs are generated in a vacuum by the application of an electric field. 
Recent research on the dynamically assisted Schwinger mechanism calculates the particle pair creation rate analytically using the Furry picture (FP)~\cite{furry1951bound} for a system in which an oscillating electric field is perturbatively added to a strong constant electric field. It has been suggested that the perturbed electric field allows non-adiabatic contributions to appear~\cite{taya2019franz,huang2019spin}.

In this paper, we consider the transverse field Ising model which depends linearly on time, with time-periodic perturbations and investigate how the addition of oscillations affects the phase transition behavior in the QKZM framework. The analytical expression of the density of defects is derived using the Landau--Zener--St\"ukelberg--Majorana (LZSM) model~\cite{landau1932theorie,zener1932non,stueckelberg1932two,majorana1932atoms}. The LZSM model describes a two-level system whose Hamiltonian has diagonal elements that are linearly dependent on time, while the off-diagonal elements are time-independent. 
The calculations are performed using the perturbation and FP formulation to derive analytical solutions with approximations. The perturbation approximation is valid for the non-adiabatic region, while the FP formulation is valid for the adiabatic region.

The structure of this paper is as follows. 
In Sec.~\ref{sec:two-level}, we analyze the contribution of time-periodic perturbations to the two-level system for the calculations in the next section. In this section, we introduce the LZSM model and analyze the dynamics of the system when time-periodic perturbations are added, using the perturbation theory and the FP formalism. Furthermore, we confirm that these approximate solutions are in good agreement with numerical calculations.
In Sec.~\ref{sec:ising}, we consider a time-dependent transverse field Ising chain with a time-periodic perturbation. We determine how the density of defects changes when a time-periodic perturbation is applied to the diagonal or off-diagonal elements, and compare the results with those obtained by the QKZM.
In Sec.~\ref{sec:conclusion}, we summarize the discussion so far.

\section{Transition Probabilities of Two-Level System}\label{sec:two-level}

This section focuses on the treatment of the LZSM model in the presence of an external oscillation field to analyze many-body systems later. There have been some previous studies on this topic~\cite{mullen1989time,kayanuma2000landau}. Here, we introduce the LZSM model first, and the perturbation theory and the FP formulation for the LZSM model. Finally, we evaluate the validity of these approximations. 

\subsection{LZSM model}

The LZSM model is described by the two-level Hamiltonian
\begin{align}
H_\mathrm{LZSM}(t)&=\frac{1}{2}vt\sigma^z+\Delta\sigma^x.
\end{align}
In this model, if a state was an instantaneous eigenstate in the infinite past, the probability of transitioning to another instantaneous eigenstate in the infinite future is given by 
\begin{align}
P_{\mathrm{LZSM}}=\exp\qty(-\frac{2\pi \Delta^2}{v}),\label{eq:lzsm_prob}
\end{align}
where the natural units are used. When $\Delta$ is significantly larger than $\sqrt{v}$, the system is considered adiabatic, resulting in a small transition probability. Conversely, when $\Delta$ is significantly smaller than $\sqrt{v}$, the system is characterized as non-adiabatic, leading to a large transition probability. 
Recent studies of the LZSM model have investigated the dynamics under various conditions, including the presence of external oscillating perturbations~\cite{mullen1989time,
malla2018landau,
wubs2005landau,
wubs2006gauging,
saito2006quantum,
saito2007dissipative,
zueco2008landau,
ashhab2014landau,
ashhab2016landau,
sinitsyn2016solvable,
sun2016landau,
kayanuma2000landau}. The perturbation approach derives the approximate formula for the LZSM model when the diagonal elements of the Hamiltonian are small~\cite{mullen1989time}.

\subsection{perturbation theory}

In this section, we consider the two-level time-dependent Hamiltonian
\begin{align}
H(t)&=H_z(t) +H_x(t), \label{eq:ham_2level}\\
H_z(t)&=\frac{1}{2}(vt+\varepsilon-A\cos(\omega t))\sigma^z,\\
H_x(t)&=\qty(\Delta+\frac{B}{2} \cos \omega t)\sigma^x,
\end{align}
which is the LZSM model with oscillations of magnitude $A$ and $B$ in the diagonal and off-diagonal elements, respectively. The initial state is assumed to be $\ket{\psi(-\infty)}\propto\ket{\uparrow}$, where $\sigma_z\ket{\uparrow}=\ket{\uparrow}$ holds. The goal is to obtain the transition probability at the final time $p(\infty)=|\bra{\psi(\infty)}\ket{\uparrow}|^2$. We note that the transition probabilities were obtained approximately when either $A$ or $B$ is $0$ with perturbation theory~\cite{mullen1989time}. Changing the frame with the unitary operator
\begin{align}
U_z(t)=\exp\qty(-i\int_0^tdt' H_z(t')),
\end{align}
the Schr\"odinger equation $i\dot U(t)=H(t)U(t)$ becomes
\begin{align}
    i\frac{d}{dt}\hat U_x(t)&=\hat H_x(t) \hat U_x(t),
\end{align}
where we define $\hat U_x(t)=U_z^\dagger(t)U(t)$ and 
\begin{widetext}
    \begin{align}
    \hat H_x(t)&=U_z^\dagger(t) H_x(t)U_z(t) \\
    &=\qty(\Delta+\frac{B}{2} \cos \omega t)\sum_{n=-\infty}^\infty\mqty(0&J_n\qty(\frac{A}{\omega})e^{\frac{i}{2}vt^2+i\varepsilon t-in\omega t}\\ J_n\qty(\frac{A}{\omega})e^{-\frac{i}{2}vt^2-i\varepsilon t+in\omega t}&0).
    \end{align}
Here, we used the formula
\begin{align}
    e^{i x \sin \tau}=\sum_{n=-\infty}^{\infty} J_n(x) e^{i n \tau},
\end{align}
where $J_n(x)$ is the Bessel function of the first kind and the basis of the matrix is $\{\ket{\uparrow},\ket{\downarrow}\}$, where $\sigma_z\ket{\downarrow}=-\ket{\downarrow}$ holds. 
We express the state in this basis as $\ket{\tilde \psi(t)}=\mqty(C_{\uparrow}(t)&C_{\downarrow}(t))^\mathrm{T}$, and these variables satisfy
\begin{align}
i  \dot C_{\uparrow}(t)&=\qty(\Delta+\frac{B}{2}\cos\omega t) \sum_{n=-\infty}^{\infty} J_n\qty(\frac{A}{\omega})e^{i \frac{v}{2} t^2 +i\varepsilon t-in\omega t} C_{\downarrow}(t),\\
i  \dot C_{\downarrow}(t)&=\qty(\Delta+\frac{B}{2}\cos\omega t) \sum_{n=-\infty}^{\infty} J_n\qty(\frac{A}{\omega})e^{-i \frac{v}{2} t^2 -i\varepsilon t+in\omega t} C_{\uparrow}(t).
\end{align}
The initial conditions on these variables can be regarded as $C_\uparrow(-\infty)=1$ and $ C_\downarrow(-\infty)=0$, and the transition probability $p(\infty)$ can be expressed as $|C_\uparrow(\infty)|^2$. 
We introduce the dimensionless parameters 
$\tau=\sqrt{v} t$ and $\eta=A/\omega$, and we define $\tilde\circ=\circ/\sqrt{v}$. By successive substitutions, we obtain the following result 
\begin{align}
C_\uparrow(\infty)-1
&\simeq -\sum_{n=-\infty}^\infty\sum_{m=-\infty}^\infty J_n(\eta)J_m(\eta)\\
&\qquad \times \int^{\infty}_{-\infty}d{\tau}\int^{\tau}_{-\infty}d{\tau'}\ \qty(\tilde\Delta+\frac{\tilde B}{2}\cos\tilde\omega \tau)\qty(\tilde\Delta+\frac{\tilde B}{2}\cos\tilde\omega \tau')
e^{\frac{i}{2}({\tau}^2-\tau'^2)+i\tilde \varepsilon (\tau-\tau')-i\tilde\omega (n \tau-m\tau')}\\
&=-2\pi
\sum_{n=-\infty}^\infty\sum_{m=-\infty}^\infty\qty(\tilde\Delta J_{n}(\eta)+\frac{\tilde B}{4}J_{n+1}(\eta)+\frac{\tilde B}{4}J_{n-1}(\eta)) \qty(\tilde\Delta J_{m}(\eta)+\frac{\tilde B}{4}J_{m+1}(\eta)+\frac{\tilde B}{4}J_{m-1}(\eta))\\
&\qquad \times \exp\qty(-\frac{i}{2}\tilde\omega(n^2\tilde\omega-2n\tilde\varepsilon)+\frac{i}{2}\tilde\omega(m^2\tilde\omega-2m\tilde\varepsilon))\theta(n-m),
\end{align}
where we define the step function
\begin{align}
\theta(x)=
\left\{
\begin{array}{l}
1 \quad (x>0)\\
1/2\quad (x=0)\\
0\quad (x<0)
\end{array}
\right. .
\end{align}
Here, we assume $\tilde \Delta$ and $\tilde B$ are small enough that this approximation is valid in the non-adiabatic region.
Finally, the transition probability is approximately
\begin{align}
p(\infty)
&\simeq \exp\Biggl(-4\pi
\sum_{n=-\infty}^{\infty}\sum_{m=-\infty}^{\infty}\qty(\tilde\Delta J_{n}(\eta)+\frac{\tilde B}{4}J_{n+1}(\eta)+\frac{\tilde B}{4}J_{n-1}(\eta)) \qty(\tilde\Delta J_{m}(\eta)+\frac{\tilde B}{4}J_{m+1}(\eta)+\frac{\tilde B}{4}J_{m-1}(\eta)) \\
&\qquad \times\cos\qty(\frac{1}{2}\tilde\omega(n^2\tilde\omega-2n\tilde\varepsilon)-\frac{1}{2}\tilde\omega(m^2\tilde\omega-2m\tilde\varepsilon))\theta(n-m)\Biggr)\\
&=:P_{\mathrm{PT}}.\label{eq:p_2level_fp}
\end{align}
\end{widetext}
We note that the transition probability $p(\infty)$ must be periodic with $\varepsilon$ because
\begin{align}
H\qty(t,\varepsilon-2\pi n \frac{v}{\omega})&=H\qty(t-n\frac{2\pi}{\omega},\varepsilon),\quad \forall n\in\mathbb{Z}
\end{align}
holds and the period is $2\pi/\tilde\omega$.

\subsection{Furry Picture}

Next, we decompose the Hamiltonian~\eqref{eq:ham_2level} as in
\begin{align}
    H(t)
    &=H_0(t)+H_1(t),\\
    H_0(t)&= \frac{1}{2}(v t+\varepsilon)\sigma_z+\Delta\sigma_x,\\
    H_1(t)&=\frac{1}{2}\cos\omega t(-A\sigma_z+B\sigma_x),
\end{align}
where $H_0(t)$ is the Hamiltonian of the LZSM model.
Let $U_0(t)$ be the time-evolution operator of $H_0(t)$ and we define $\hat H_1(t)=U_0^\dagger(t) H_1(t)U_0(t)$.
If the $\hat H_1(t)$ is sufficiently small, we can approximate the time-evolution operator by first-order:
\begin{align}
    U(t)&\simeq U_0(t)\qty( I-i\int_{t_0}^t dt' \, \hat H_1(t')).
\end{align}
Then, the transition probability becomes
\begin{align}
    p(\infty)&\simeq \qty|\bra{\uparrow}U_0(\infty)\ket{\uparrow}-i\bra{\uparrow}U_0(\infty)\int_{-\infty}^\infty dt' \, \hat H_1(t')\ket{\uparrow}|^2 \\
    &=:P_{\textrm{FP}}.
\end{align}
We note that we need to consider up to the second-order perturbation if we approximate the transition probability by the second-order of $\hat H_1$:
\begin{align}
p(\infty)&\simeq  P_{\textrm{FP}}  -2 \operatorname{Re}\biggl(\left\langle \uparrow\left|U_0(\infty)\right| \uparrow\right\rangle^*\\
&\quad \times \langle \uparrow|U_0(\infty) \int_{-\infty}^{\infty} d t \int_{-\infty}^t d t^{\prime} \hat{H}_1(t) \hat{H}_1\left(t^{\prime}\right)|\uparrow\rangle\biggr).
\end{align}
However, we assume that the last term is negligible. This assumption is justified in the adiabatic limit because the term contains an exponentially small term in the limit.
The above method is called the Furry picture formalism~\cite{furry1951bound}.

With this formalism, the transition probability in the adiabatic limit is
\begin{align}
P_{\mathrm{FP}}
&\simeq \left|\bra{\downarrow}\int_{-\infty}^{\infty} d t \hat{H}_1(t)\ket{\uparrow}\right|^2 \\
&\simeq  \pi^2e^{-2\pi\tilde\Delta^2}\biggr|\eta {}_1\tilde F_1\qty(-i\tilde\Delta^2,0,i\tilde\omega^2)\\
&\quad -i\tilde B \frac{|\tilde\Delta|}{2}\biggl({}_1\tilde{F}_1\left(-i \tilde\Delta^2,1,i\tilde\omega^2\right)\\
&\quad +{}_1\tilde{F}_1\left(-i \tilde\Delta^2+1,1,i\tilde\omega^2\right)\biggr)\biggr|^2\label{eq:p_2level_fk},
\end{align}
where ${}_1\tilde F_1(a,b,x)$ is the regularized confluent hypergeometric function of the first kind. We derive the probability in Appendix.\ref{appsec:FP} and the  probability without the adiabatic limit is \eqref{eq:prob_fk_exact}.
The first-order approximation in this method is valid in the region where both $\eta,\tilde B$ are small. Unlike perturbation theory, this method does not treat $\tilde\Delta$ as small values, but rather assumes that it takes on large values. In this way, this method calculates non-perturbative effects on $\tilde \Delta$.

\subsection{numerical calculation}

In this subsection, we compare the approximate formula~\eqref{eq:p_2level_fp} and~\eqref{eq:prob_fk_exact} or~\eqref{eq:p_2level_fk} with the results of numerical solution of the Schr\"odinger equation. 

First, consider the case of the non-adiabatic limit and the case $\tilde A=0$. In this case, only the off-diagonal component has an oscillating term and the transition probability~\eqref{eq:p_2level_fp} can be expressed as 
\begin{align}
    P_{\mathrm{PT}}&= \exp\biggl(-2\pi\tilde\Delta^2-\frac{\pi}{2}
\tilde B^2 \cos^2\qty(\tilde\omega\tilde\varepsilon)\\
&\qquad -2\pi
\tilde B\tilde\Delta \cos\qty(\frac{\tilde\omega^2}{2})\cos\qty(\tilde\omega\tilde\varepsilon)\biggr)\label{eq:p_2level_a0}.
\end{align}
The result~\eqref{eq:p_2level_a0} is also derived in the previous study \cite{mullen1989time}. We note, however, that a factor of $1/4$ is missing in the third term of equation (7) in~\cite{mullen1989time}.

\begin{figure}[!t]
    \centering
        \begin{overpic}[width=0.9\linewidth]{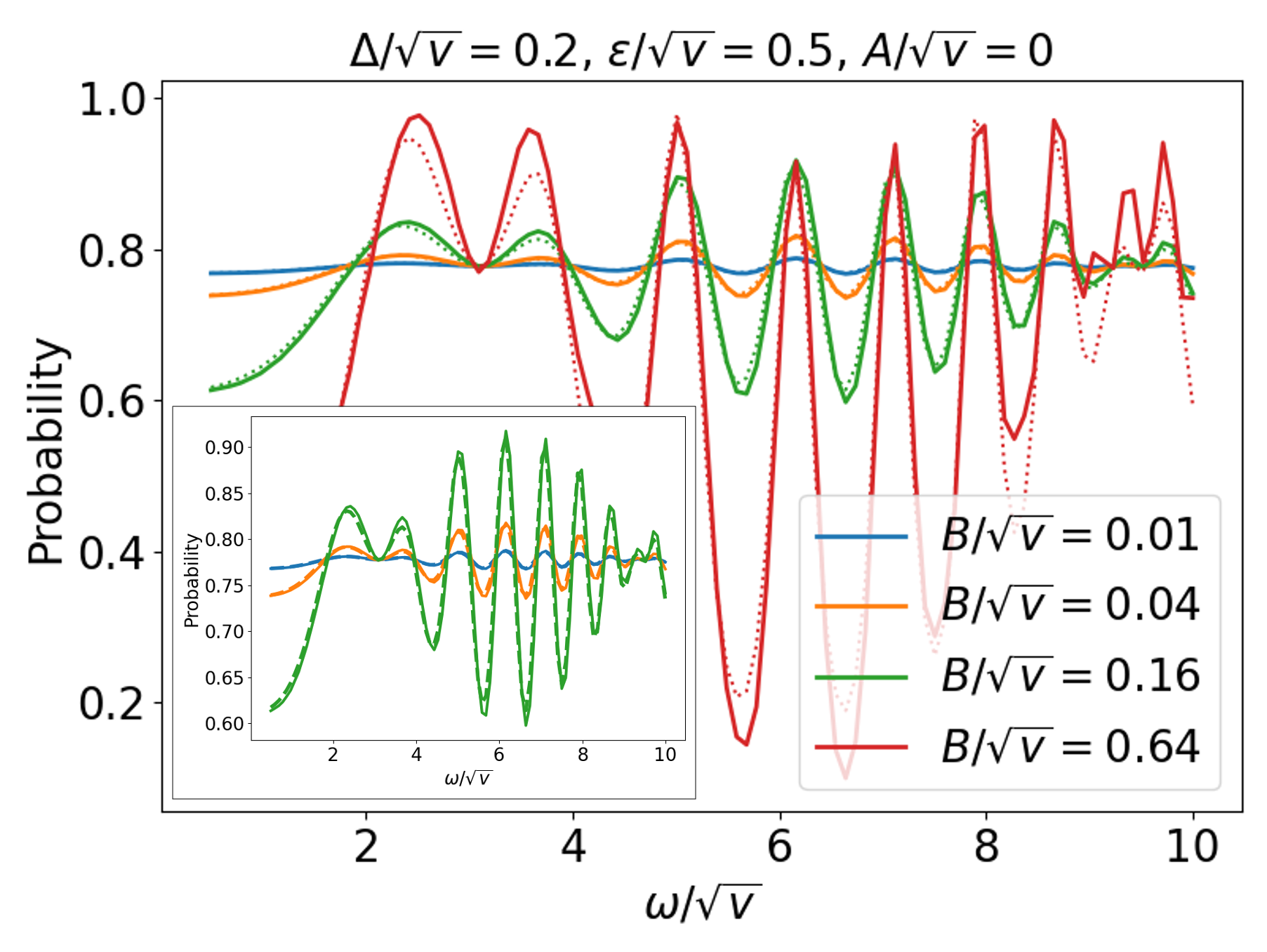}
        \put(12,73){\large{(a)}}
        \end{overpic}
        \begin{overpic}[width=0.9\linewidth]{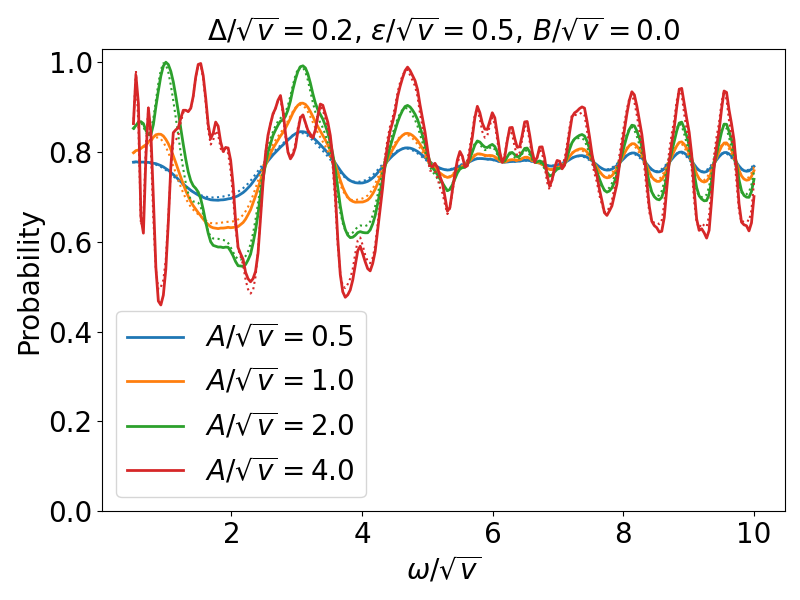}
        \put(12,73){\large{(b)}}
        \end{overpic}
        \begin{overpic}[width=0.9\linewidth]{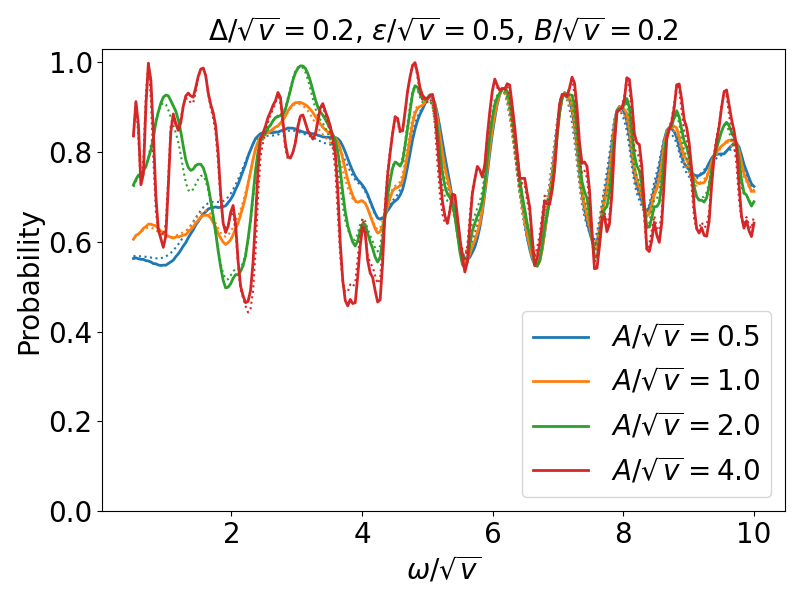}
        \put(12,73){\large{(c)}}
        \end{overpic}
        \caption{Numerical calculations (solid) and analytical approximate solutions (dotted) are plotted. Inset is a magnified view of the vertical axis. Parameters are (a) $\tilde\Delta=0.2,\ \tilde\varepsilon=0.5,\  \tilde A=0$, (b) $\tilde \Delta=0.75,\  \tilde\varepsilon=0.5,\  \tilde B=0$, and (c) $\tilde \Delta=0.2,\   \tilde\varepsilon=0.5,\  \tilde B=0.2$. In (b) and (c), the sum was calculated in the range of $-10\leq n,m \leq 10$. We can see that the numerical calculation and the approximate formulae agree well in the region where $\tilde B$ is small. 
        }
        \label{fig:prob_fp}
\end{figure}

The numerical results in this case are shown in Fig.~\ref{fig:prob_fp}(a). In the region where $\tilde B$ is small, the numerical calculation and the approximate expression~\eqref{eq:p_2level_a0} are in good agreement. On the other hand, as $\tilde B$ increases, the contribution of $O(\tilde B^3)$, which is ignored in the approximate expression~\eqref{eq:p_2level_a0}, increases, resulting in deviation from the numerical calculation.

Next, consider the case $\tilde B=0$. From \eqref{eq:p_2level_fp}, the transition probability can be expressed as 
\begin{align}
    P_{\mathrm{PT}}&= \exp\biggl(-4\pi\tilde\Delta^2 
\sum_{n=-\infty}^{\infty}\sum_{m=-\infty}^{\infty}J_{n}(\eta) J_{m}(\eta)\\
& \times \cos\qty(\tilde\omega\qty(\frac{1}{2}(n^2-m^2)\tilde\omega-(n-m)\tilde\varepsilon))\theta(n-m)\biggr)\label{eq:p_2level_b0}.
\end{align}

The numerical results in this case are shown in Fig.~\ref{fig:prob_fp}(b). Here, the sum was calculated in the range of $-10\leq n,m \leq 10$. In this case, the sum is large enough that even if $\eta=A/\omega$ is not small, the numerical calculation and the approximate expression~\eqref{eq:p_2level_b0} are in good agreement. In Fig.~\ref{fig:prob_fp}(b), the transition probabilities show a simple behavior as $\tilde\omega$ increases. This corresponds to the region where $\eta$ is sufficiently small. In this limit, \eqref{eq:p_2level_b0} yields
\begin{align}
    P_{\mathrm{PT}}&\simeq \exp\biggl(-2\pi\tilde\Delta^2 
\biggl(1+2\eta\sin(\tilde\omega\tilde\varepsilon)\sin\frac{\tilde\omega^2}{2}\biggr)\biggr).
\end{align}
This result is also derived in the previous study \cite{mullen1989time}.

Finally, consider the case $\tilde A\neq 0$ and $\tilde B\neq 0$. The numerical results for this case are shown in Fig.~\ref{fig:prob_fp}(c). In this case, the numerical calculation and the approximate formula~\eqref{eq:p_2level_fp} agree well even for large values of $\eta$ because the sums are sufficiently large.

\begin{figure}[!t]
    \centering
        \begin{overpic}[width=0.9\linewidth]{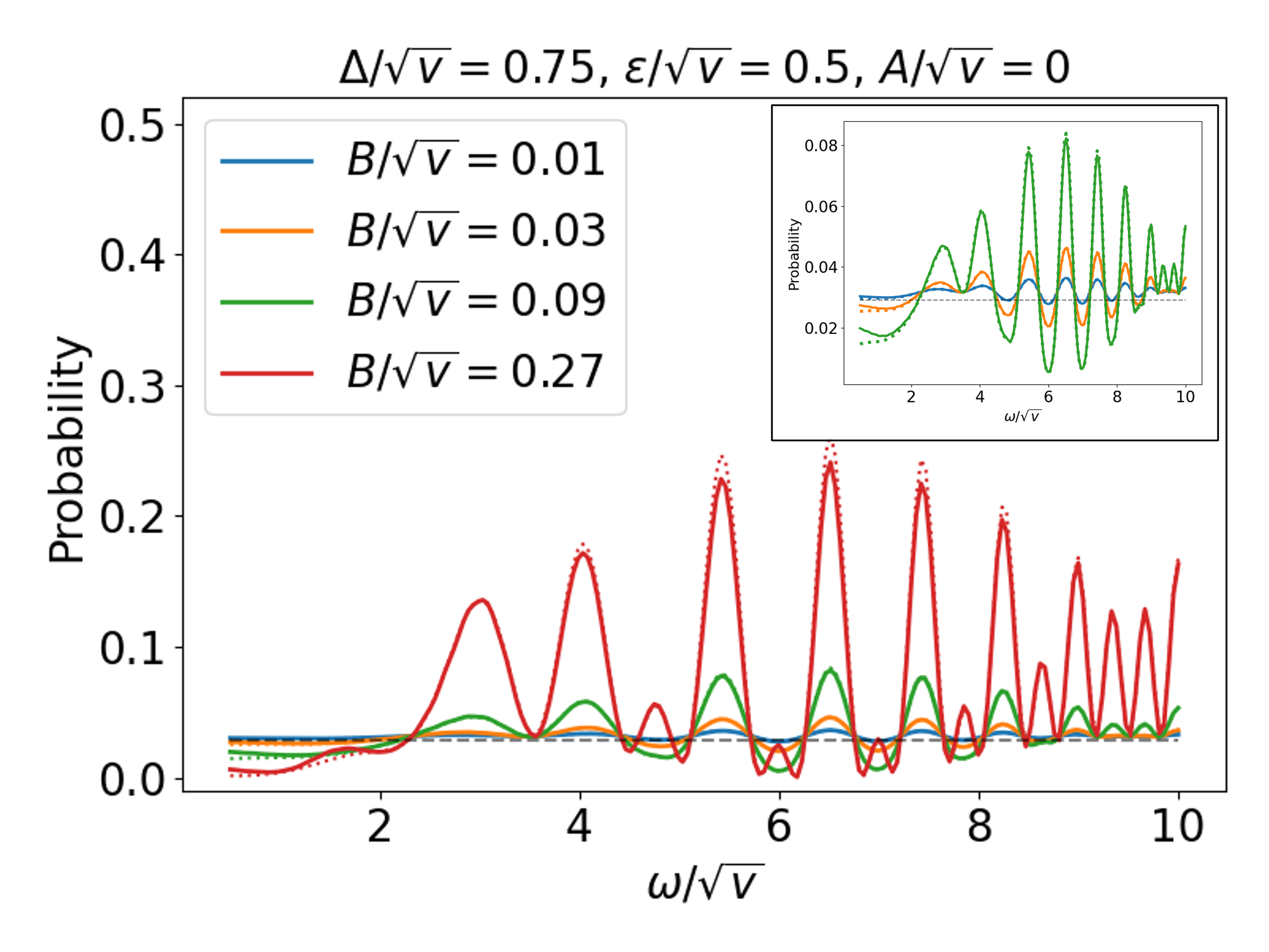}
        \put(12,73){\large{(a)}}
        \end{overpic}
        \begin{overpic}[width=0.9\linewidth]{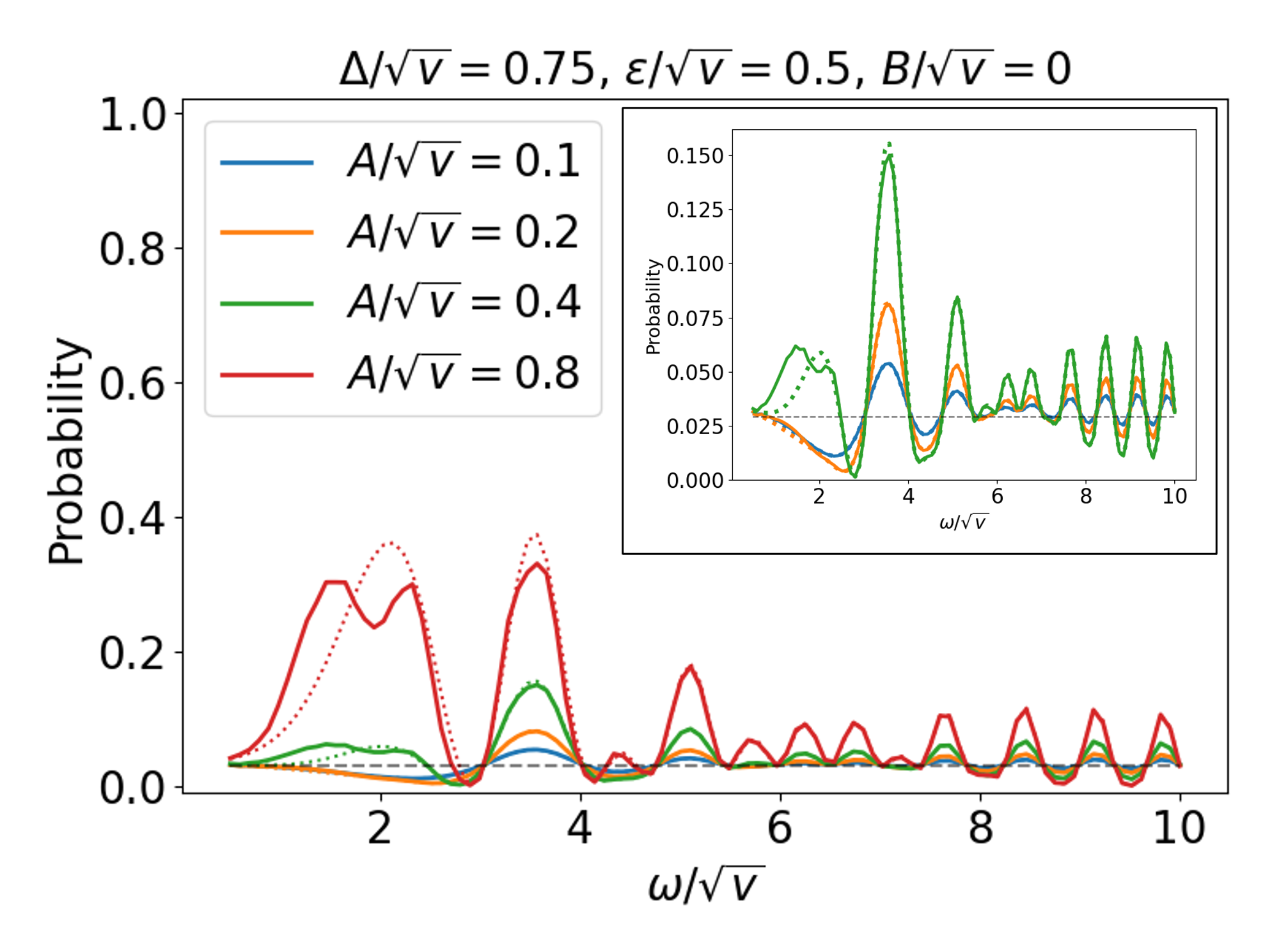}
        \put(12,73){\large{(b)}}
        \end{overpic}
        \begin{overpic}[width=0.9\linewidth]{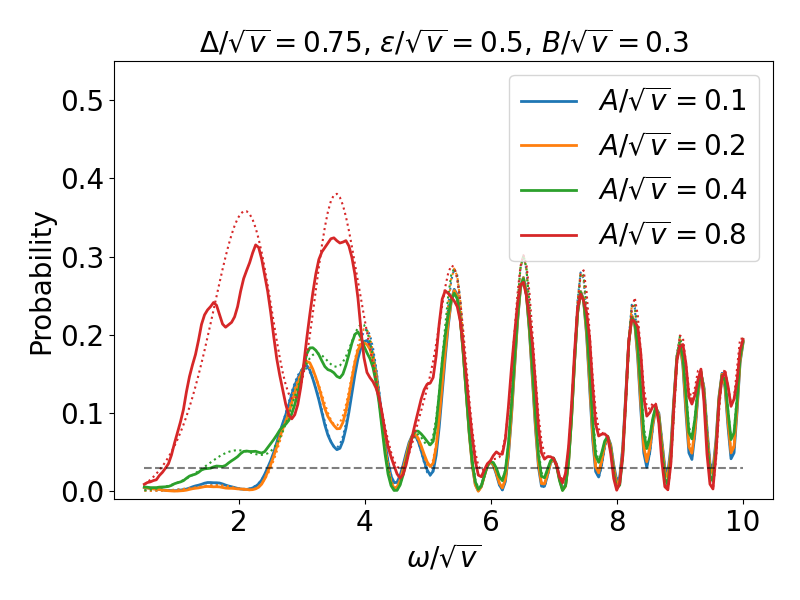}
        \put(12,73){\large{(c)}}
        \end{overpic}
        \caption{Numerical calculations (solid), analytical approximate solutions (dotted), and LZSM transition probability $P_{LZ}$ (dashed) are plotted. Inset is a magnified view of the vertical axis. Parameters are (a) $\tilde\Delta=0.75, \ \tilde\varepsilon=0.5,\  \tilde A=0$, (b) $\tilde \Delta=0.75,\  \tilde\varepsilon=0.5,\  \tilde B=0$, and (c) $\tilde \Delta=0.75,\   \tilde\varepsilon=0.5,\  \tilde B=0.3$. We can see that the numerical calculation and the approximate formula agree well in the region where $\eta$ and $\tilde B$ are small. 
        }
        \label{fig:prob_fk}
\end{figure}

Next, we show the validity of \eqref{eq:prob_fk_exact} in the adiabatic process. First, in the case of $\tilde A=0$, the results of the numerical calculations are compared with those of the expression \eqref{eq:prob_fk_exact} in Fig.~\ref{fig:prob_fk}(a). It can be seen that in the region where $\tilde B$ is small, the results are in good agreement with the numerical calculations.
The dashed line in the figure represents the LZSM transition probability~\eqref{eq:lzsm_prob}. Although this probability is sufficiently small in the adiabatic limit, it can be seen that there are parameter regions where the transition probabilities are much larger than the LZSM transition probability for $\tilde B\neq0$ due to the effects of the oscillations.

Next, in the case of $\tilde B=0$, the results of the numerical calculations are compared with those of \eqref{eq:prob_fk_exact} in Fig.~\ref{fig:prob_fk}(b). It can be seen that in the region where $\eta$ is sufficiently small, the results are in good agreement with the numerical calculations. In this case, as in the previous case, there are parameter regions where the transition probabilities are much larger than the LZSM transition probability due to the oscillations.

In addition, Fig.~\ref{fig:prob_fk}(c) compares the results of the numerical calculations with those of \eqref{eq:prob_fk_exact} in the case of $A\neq 0$ and $B\neq 0$. In this case, we can see that \eqref{eq:prob_fk_exact} is in good agreement with the numerical calculation in the region where $\eta$ is sufficiently small due to the small value of $\tilde B$.

\begin{figure}[t!]
    \centering
    \includegraphics[width=0.9\linewidth]{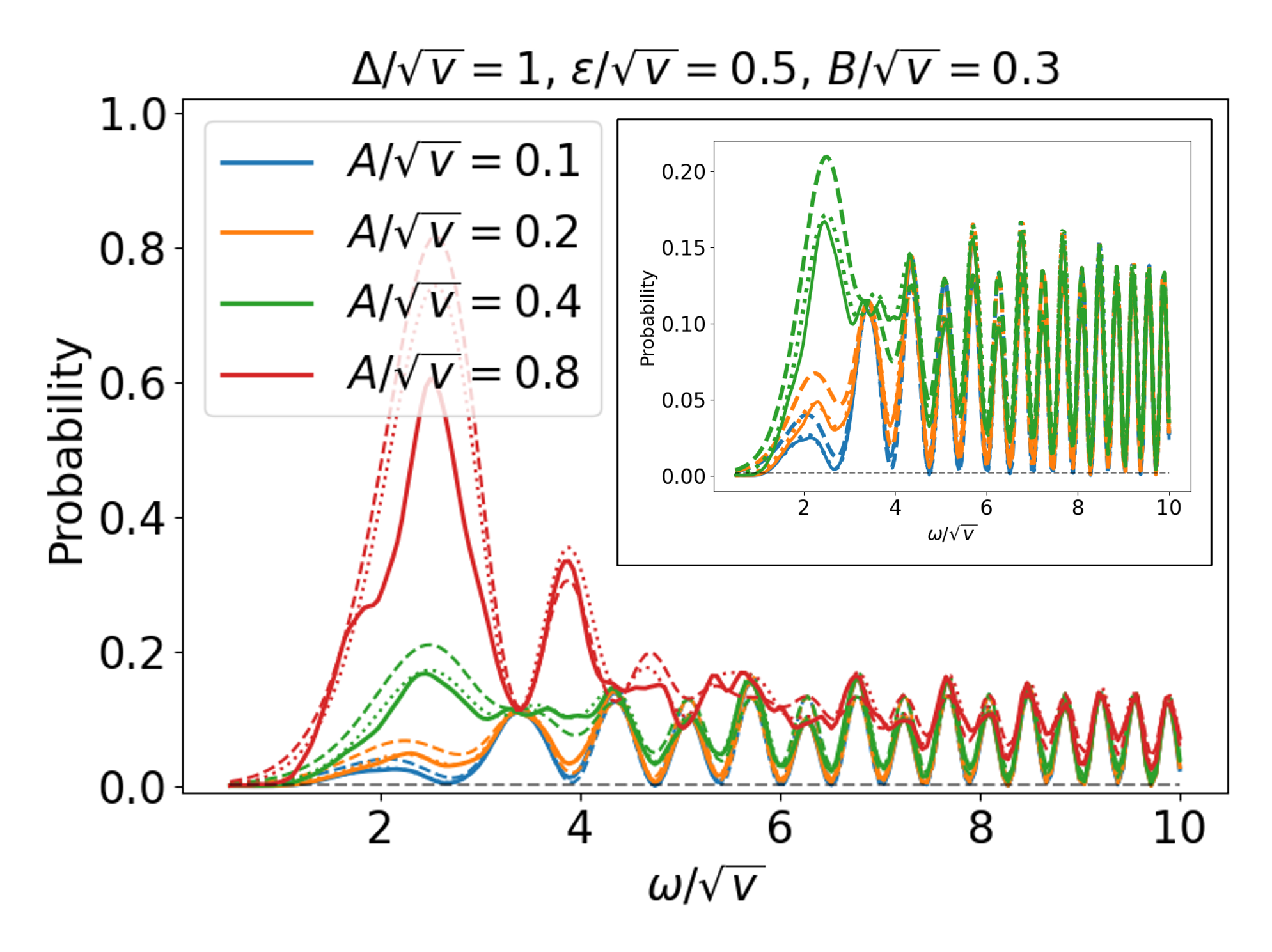}
    \caption{Numerical calculations (solid), analytical approximate solutions (dotted,\eqref{eq:prob_fk_exact}), and analytical approximate solutions in the adiabatic limit (dashed,\eqref{eq:p_2level_fk}) are plotted. Inset is a magnified view of the vertical axis. Parameters are $\tilde \Delta=1.0, \ \tilde\varepsilon=0.5,\  \tilde B=0.3$. It can be seen that the dotted lines \eqref{eq:prob_fk_exact} and the dashed lines \eqref{eq:p_2level_fk} agree well, especially in regions where $\eta$ is sufficiently small.
    }
    \label{fig:prob_lz_fk_adi}
\end{figure}

Finally, we check that \eqref{eq:p_2level_fk} is consistent with \eqref{eq:prob_fk_exact} in the adiabatic limit. The results are shown in Fig.~\ref{fig:prob_lz_fk_adi}. In this case, it can be seen that \eqref{eq:prob_fk_exact} is consistent with \eqref{eq:p_2level_fk}, especially in regions where $\eta$ is sufficiently small.

\section{TRANSVERSE ISING CHAIN with time-periodic perturbation}\label{sec:ising}

Next, we consider the transverse field Ising model which depends linearly on time, with time-periodic perturbations. For this model, there is a previous study that investigated the model numerically~\cite{mukherjee2009effects}. However, this study only shows that the transfer matrix method~\cite{kayanuma2000landau} agrees with the numerical calculations. In the following, we consider the case where the perturbations are uniformly contained in the diagonal or off-diagonal elements.

\subsection{perturbation in the diagonal elements}\label{diag}

\begin{figure}[t]
    \centering
    \includegraphics[width=0.95\linewidth]{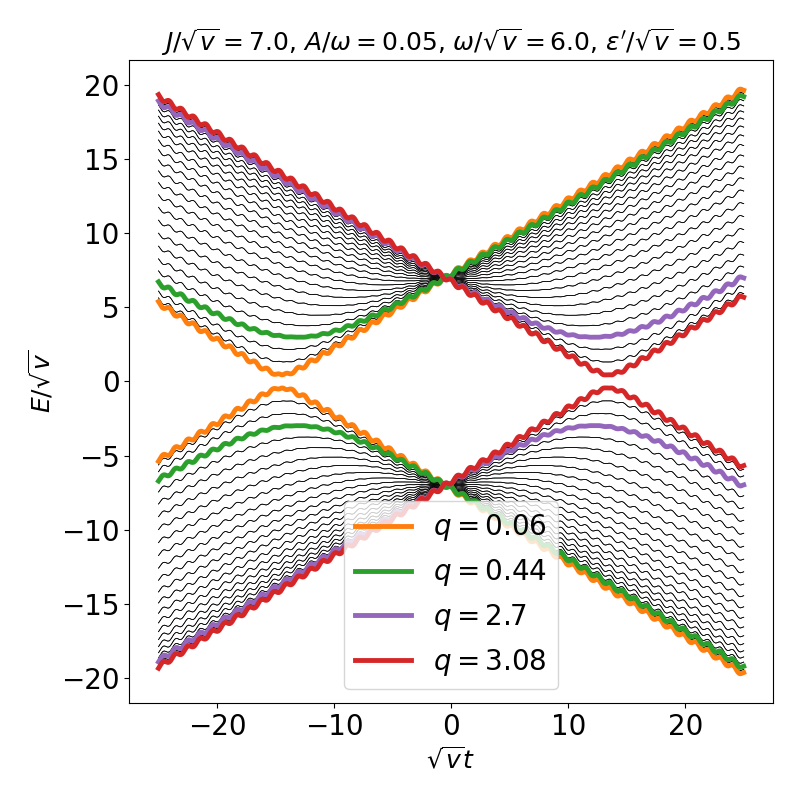}
    \caption{Time dependence of instantaneous eigenvalues of the Hamiltonian in \eqref{eq:sch_cq} for $\tilde J=7.0,\ \eta =0.05,\ \tilde\omega =6.0,\ \tilde\varepsilon'=0.5$. It is plotted as $N=50$. The two-level system represented by the line (orange, red) corresponds to a non-adiabatic region with a narrow gap, while the line (green, purple) corresponds to an adiabatic region with a wide gap.}
    \label{fig:energy_kz_N200_eta25_tomega25}
\end{figure}

We consider the time-dependent Hamiltonian
\begin{align}
H_D(t)&=-\sum^{N}_{j=1}\qty(\frac{J}{2}\sigma^x_{j}\sigma^x_{j+1}+g(t)\sigma^z_{j})\label{eq:hamiltonian},\\
g(t)&=\frac{1}{4}\qty(vt+\varepsilon'-A\cos(\omega t)),
\end{align}
where we impose the periodic boundary condition
\begin{align}
\sigma^a_{N+j}=\sigma^a_{j}.
\end{align}
This Hamiltonian has $\mathbb{Z}_2$ symmetry and only the space to which the ground state belongs will be considered from now on. Here, we introduce the spinless fermion operators $c_j$ using the Jordan--Wigner(JW) transformation
\begin{align}
\sigma^z_j=1-2c^{\dagger}_j c_j,\quad \sigma^x_j=\qty(c^{\dagger}_j+c_j)\prod_{l<j}\qty(-\sigma^z_l),
\end{align}
and we consider the Fourier expansion of the operators
\begin{align}
c_j=\frac{1}{\sqrt{N}} e^{-i\frac{\pi}{4}} \sum_q e^{iqj}c_q,
\end{align}
where $q=\pm (2 n-1) \pi/N, \ n\in\{1, \cdots,N/2\}$. In the Heisenberg picture, these operators satisfy 
\begin{align}
    i\frac{d}{dt}\mqty(c_q(t)\\c^\dagger_{-q}(t))&=\mqty( E_q(t)&\delta_q\\\delta_q&- E_q(t))\mqty(c_q(t)\\c^\dagger_{-q}(t))\label{eq:sch_cq},\\
     E_q(t)&=J\cos q+\frac{1}{2}\qty(vt+\varepsilon'-A\cos(\omega t)),\\
    \delta_q&=-J\sin q.
\end{align}
The eigenvalues of the Hamiltonian in \eqref{eq:sch_cq} are shown in Fig.~\ref{fig:energy_kz_N200_eta25_tomega25}. It can be seen that when $q\simeq 0,\pm\pi$, the energy gap is small, corresponding to the non-adiabatic region where non-adiabatic transitions occur, while in the other region, the energy gap is large, corresponding to the adiabatic region.

The initial state is set to be the ground state at $t=-\infty$ : $\ket{\psi(-\infty)}=\ket{\downarrow}^{\otimes N}$. The final time is set to $t=t_F$ and we calculate the expectation value
\begin{align}
    \expval{\mathcal{N}}&=\frac{1}{2}\bra{\psi(-\infty)}\sum_j\qty(1-\sigma^z_j(t_F))\ket{\psi(-\infty)}\\
    &=\sum_{j=1}^N\bra{\psi(-\infty)}c^{\dagger}_j (t_F)c_j(t_F)\ket{\psi(-\infty)}\\
    &=\sum_{q} \bra{1(q)}  c_q^{\dagger}(t_F)  c_q(t_F)\ket{1(q)},
\end{align}
where $c^\dagger_q(-\infty) c_q(-\infty)\ket{1(q)}=\ket{1(q)}$ and we used the fact that the initial state can be written as $\bigotimes_{q}\ket{1(q)}$. The solution of \eqref{eq:sch_cq} can be expressed as
\begin{align}
    \mqty(c_q(t)\\c_{-q}^\dagger(t))&=\begin{pmatrix}
    u_q(t)&-v^*_{q}(t)\\
    v_q(t)&u^*_q(t)
    \end{pmatrix}
    \mqty(c_q(-\infty)\\c_{-q}^\dagger(-\infty)),
\end{align}
where $|u_q(t)|^2+|v_q(t)|^2=1$ holds. Then, the expectation value becomes
\begin{align}
    \expval{\mathcal{N}}
    &=\sum_{q} \bra{1(q))}  c_q^{\dagger}(t_F)  c_q(t_F)\ket{1(q)}\\
    &=\sum_{q}   \qty|u_q(t_F)|^2.
\end{align}
In the thermodynamic limit, the normalized expectation value can be expressed as
\begin{align}
n(t_F)&=\frac{\expval{\mathcal{N}(t_F)}}{N}
\\
&\to \int^{\pi}_{-\pi}\frac{dq}{2\pi} \qty|u_q(t_F)|^2.
\end{align}

In the model considered in this study, the phase transition points exist at times satisfying $g(t)=\pm J/2$. However, as shown in \eqref{eq:sch_cq}, these phase transition points do not produce interference effects as discussed in previous studies~\cite{kou2022interferometry}. Therefore, we will assume that $t_F=\infty$ for the current discussion.

First, we consider a non-adiabatic region. The non-adiabatic region corresponds to the situation $\kappa_q\ll1$, where $\kappa_q=\tilde J^2\sin^2 q$. In this region, we obtain from \eqref{eq:p_2level_b0}
\begin{widetext}
\begin{align}
    |u_q(\infty)|^2&\simeq \exp\biggl(-4\pi\kappa_q 
\sum_{n=-\infty}^{\infty}\sum_{m=-\infty}^{\infty}J_{n}\qty(\eta) J_{m}\qty(\eta)\cos\qty(\tilde\omega \qty(\frac{1}{2}(n^2-m^2)\tilde\omega-(n-m)(\tilde\varepsilon'+2\tilde J\cos q))) \theta(n-m)\biggr),\label{eq:nonadi_diag}
\end{align}
where we define $\eta=A/\omega$. In the adiabatic limit where $\tilde J$ is sufficiently large, there is a non-adiabatic region only near $q=0,\pm\pi$.
In the vicinity of $q= 0$, we obtain
\begin{align}
\qty|u_q(\infty)|^2
&\simeq \exp\qty(-2\pi\tilde J^2q^2-4\pi\tilde J^2q^2\sum_{n>m}J_n\qty(\eta)J_m\qty(\eta) \cos\qty(\tilde\omega \qty(\frac{1}{2}(n^2-m^2)\tilde\omega-(n-m)(\tilde\varepsilon'+2\tilde J))))\\
&=e^{-\alpha q^2},
\end{align}
and in the vicinity of $q= \pm\pi$, we get
\begin{align}
\qty|u_q(\infty)|^2
&\simeq \exp\biggl(-2\pi\tilde J^2(q\mp \pi)^2 -4\pi\tilde J^2(q\mp \pi)^2\sum_{n>m}J_n\qty(\eta)J_m\qty(\eta) \cos\qty(\tilde\omega \qty(\frac{1}{2}(n^2-m^2)\tilde\omega-(n-m)(\tilde\varepsilon'-2\tilde J)))\biggr)\\
&=e^{-\beta (q\mp\pi)^2}.
\end{align}
\end{widetext}
We note that $|u_q(\infty)|^2$ has the finite value near $q=0,\pm\pi$ in this region if $\alpha,\beta\propto \tilde J^2$ is large enough.

Next, in the adiabatic region $\kappa_q\gg1$, we obtain
\begin{align}
     |u_q(\infty)|^2&\simeq \pi^2 \eta^2 e^{-2\pi \kappa_q} \biggl|{}_1\tilde F_1\qty(-i\kappa_q;0;i\tilde\omega^2)\biggr|^2 \\
     &=:P_{\mathrm{FP}}(q)\label{eq:ising_fk}
\end{align}
from \eqref{eq:p_2level_fk}. 

\begin{figure}[h]
    \centering
    \includegraphics[width=0.95\linewidth]{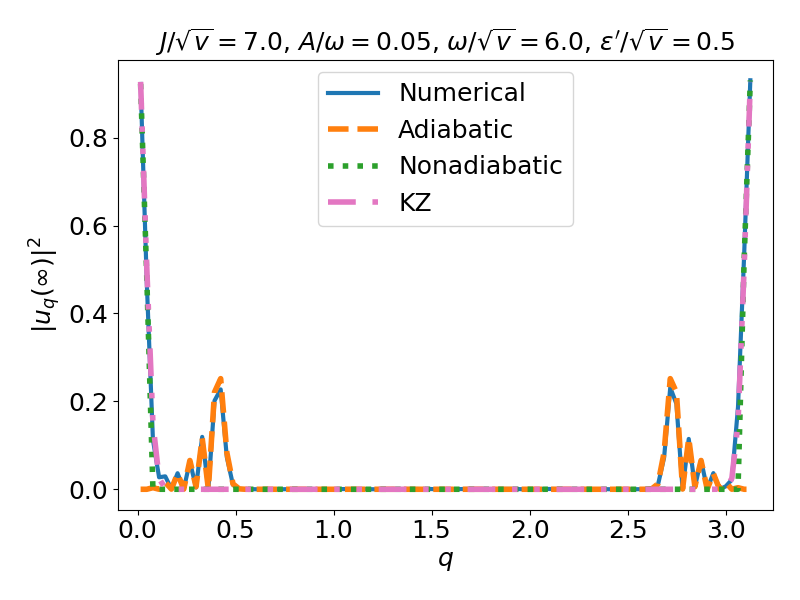}
    \caption{Comparison of numerical calculations and approximate expressions of $|u_q(\infty)|^2$ for $\tilde J=7.0,\ \eta =0.05,\  \tilde\omega=6.0,\ \tilde\varepsilon'=0.5$. It is plotted as $N=200$. The solid line shows the numerical calculation of $|u_q(\infty)|^2$, while the green dotted line is the result of plotting the approximate expression in the non-adiabatic region~\eqref{eq:nonadi_diag}. The pink dash-dot line shows the result without oscillation: $A=0$. The instantaneous eigenvalues corresponding to the maximum value in~\eqref{eq:nonadi_diag} are represented by orange and red lines in Fig.~\ref{fig:energy_kz_N200_eta25_tomega25}. The orange dashed line is the result of plotting the approximate expression in the adiabatic region~\eqref{eq:ising_fk}. The instantaneous eigenvalues corresponding to the maximum value in~\eqref{eq:ising_fk} are represented by green and purple lines in Fig.~\ref{fig:energy_kz_N200_eta25_tomega25}. }
    \label{fig:uq_kz_N_eta25_tomega25}
\end{figure}

The distribution of $|u_q(\infty)|^2$ is shown in Fig.~\ref{fig:uq_kz_N_eta25_tomega25}.
This figure shows that in addition to the transitions at $q=0,\pi$ predicted by the KZ mechanism, other transitions occur around them, which is the result from the time-periodic perturbation. 
We note that if $\tilde J$ is larger, the transitions in the adiabatic region occur closer to these vicinities as long as $\tilde\omega$ and $\eta$ are fixed.

From the above discussion, we obtain the expectation value approximately
\begin{align}
    n(\infty)&\simeq\int_{-\infty}^\infty \frac{dq}{2\pi}\qty(e^{-\alpha q^2} +e^{-\beta q^2} ) +\int_{-\pi}^\pi \frac{dq}{2\pi}P_{\mathrm{FP}}(q)\\
    &=\frac{1}{2\sqrt{\pi}}\qty(\frac{1}{\sqrt{\alpha}}+\frac{1}{\sqrt{\beta}})+ n_{\mathrm{FP}},\label{eq:diag_result}\\
    n_{\mathrm{FP}}&=\int_{-\pi}^\pi \frac{dq}{2\pi}P_{\mathrm{FP}}(q)\label{eq:n_fp}
\end{align}

\begin{figure}[h]
    \centering
    \includegraphics[width=0.95\linewidth]{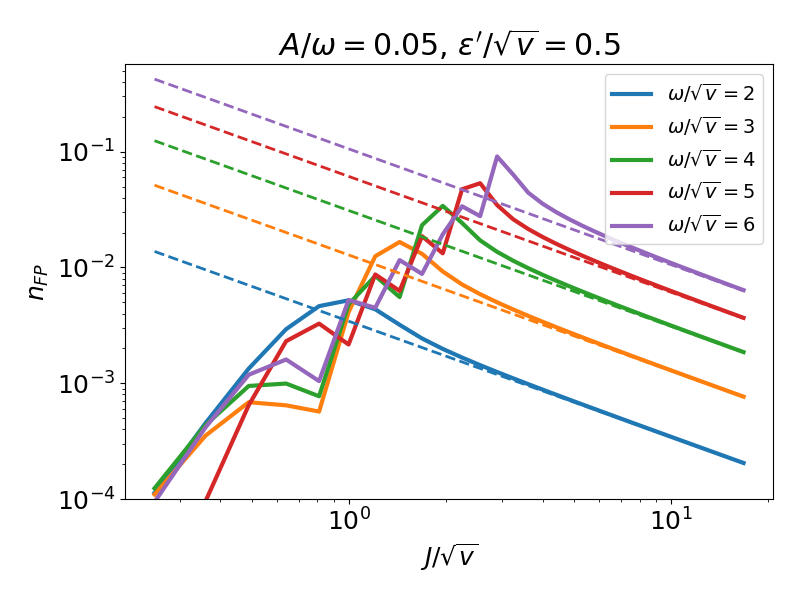}
    \caption{
    Comparison of the numerical calculation of \eqref{eq:n_fp} (solid) and the approximate expression (dashed) for $\eta =0.05, \ \tilde\varepsilon'=0.5$. The dashed lines represent \eqref{eq:n_fp_approx}. It can be seen that \eqref{eq:n_fp_approx} is an approximation where $\tilde J$ is sufficiently large and agrees well with the numerical calculation. This shows that the spin number density due to the non-perturbative effect $n_{\mathrm{FP}}$ also scales with $\tilde J^{-1}$.}
    \label{fig:prob_kz_Nintadi_eta10_tomega50_epsilon5}
\end{figure}

\begin{figure}[h]
    \centering
    \includegraphics[width=0.95\linewidth]{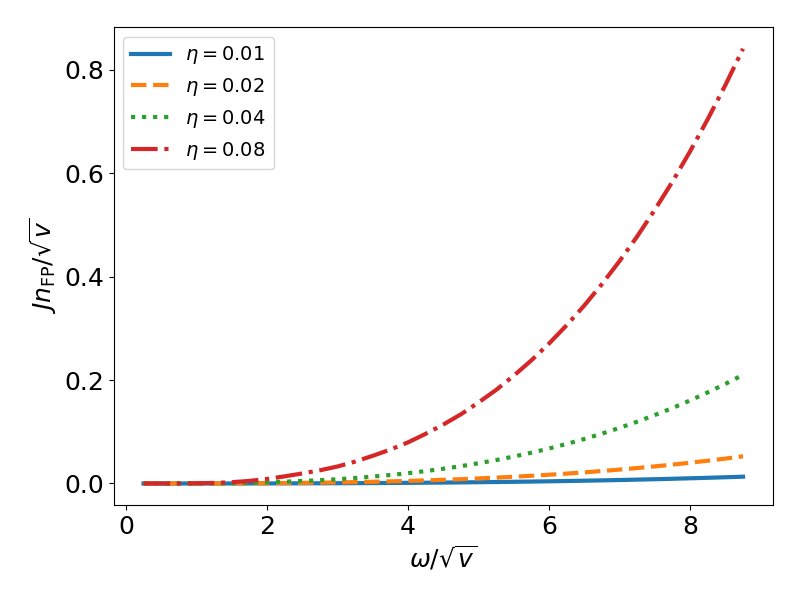}
    \caption{The dependence of the coefficient of $\tilde J^{-1}$ in~\eqref{eq:n_fp_approx} on $\tilde \omega$. The coefficient increases as $\tilde\omega$ and $\eta$ increases. This means that the effect of perturbative oscillation becomes dominant as $\tilde\omega$ and $\eta$ increases.}
    \label{fig:c_kz_Nintadi_eta10_tomega50_epsilon5}
\end{figure}

From \eqref{eq:diag_result}, we see that the first term is proportional to $\tilde J^{-1}$. This corresponds to the part of the QKZM where perturbative oscillatory effects are added to the non-adiabatic transition. The second term $n_{\mathrm{FP}}$ is the contribution from the non-perturbative effect. In fact, this term is also proportional to $\tilde J^{-1}$. 
This can be seen as follows. First, for sufficiently large $\tilde J$, $P_{\mathrm{FP}}(q)$ has a value of only $q\simeq 0, \pm\pi$, so $n_{\mathrm{FP}}$ can be transformed to 
\begin{align}
    n_{\mathrm{FP}}
    &\simeq 2\pi\eta^2\int_{0}^{\pi/2} dq \ e^{-2\pi \tilde J^2 q^2} \qty|_1\tilde{F}_1\qty(-i\tilde J^2 q^2,0,i\tilde\omega^2)|^2.
\end{align}
Transforming to $\tilde J q=x$ and changing the upper bound of the integral to $\infty$ because the integrand function has no value at $q=\pi/2$, we obtain 
\begin{align}
    n_{\mathrm{FP}}&\simeq \frac{2\pi\eta^2}{\tilde J} \int^{\infty}_{0} dx\ e^{-2\pi x^2} \qty|_1\tilde{F}_1\qty(-ix^2,0,i\tilde\omega^2)|^2.\label{eq:n_fp_approx}
\end{align}
This approximation and the original definition~\eqref{eq:n_fp} are plotted in Fig.~\ref{fig:prob_kz_Nintadi_eta10_tomega50_epsilon5}. The fact that both agree where $\tilde J$ is large indicates that this approximation is correct and that the non-perturbative contribution also shows a scaling of $\tilde J^{-1}$. It can also be seen numerically that the peak of $n_{\mathrm{FP}}$ appears where $\tilde \omega=2\tilde J$ is satisfied. This can be interpreted as a result of the resonance phenomenon.
Fig.~\ref{fig:c_kz_Nintadi_eta10_tomega50_epsilon5} shows the dependence of the coefficient of $\tilde J^{-1}$ in~\eqref{eq:n_fp_approx} on $\tilde \omega$. It can be seen that the contribution of this non-perturbative term increases as the frequency and the amplitude increases.

\begin{figure}[h]
    \centering
    \includegraphics[width=0.95\linewidth]{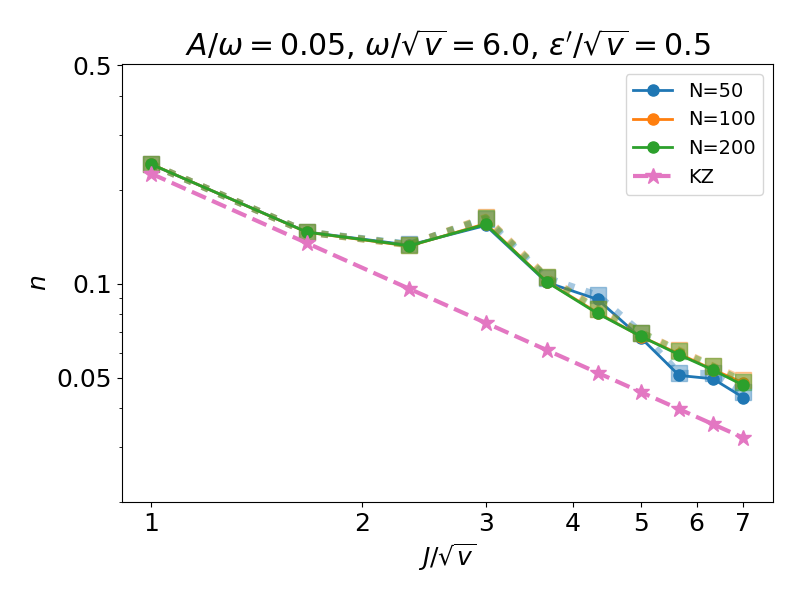}
    \caption{
    Comparison of numerical calculations and approximate expressions of the density of defects in case $N\in\{50,100,200\},\ \eta =0.05,\ \tilde\omega =6.0,\ \tilde\varepsilon'=0.5$. The solid line is the result of solving the Schr\"odinger equation numerically, the dotted line (square) corresponds to the approximate expression~\eqref{eq:diag_result}, and the pink dashed line corresponds to the result of the QKZM which is the same in $A=0$. When solving the Schr\"odinger equation numerically, the initial and final times are set to $\tau=-500,500$, respectively. It can be seen that the numerical and approximate results are in good agreement. 
    Moreover, even when $N$ is sufficiently large, the density of defects differs from that without oscillation. However, with or without oscillation, both are found to be scaled by $\tilde J^{-1}\propto \sqrt{v}$. This means that the QKZM is robust to the time-periodic perturbation.}
    \label{fig:prob_kz_N_eta25_tomega25}
\end{figure}

\begin{figure}[h]
    \centering
    \includegraphics[width=0.95\linewidth]{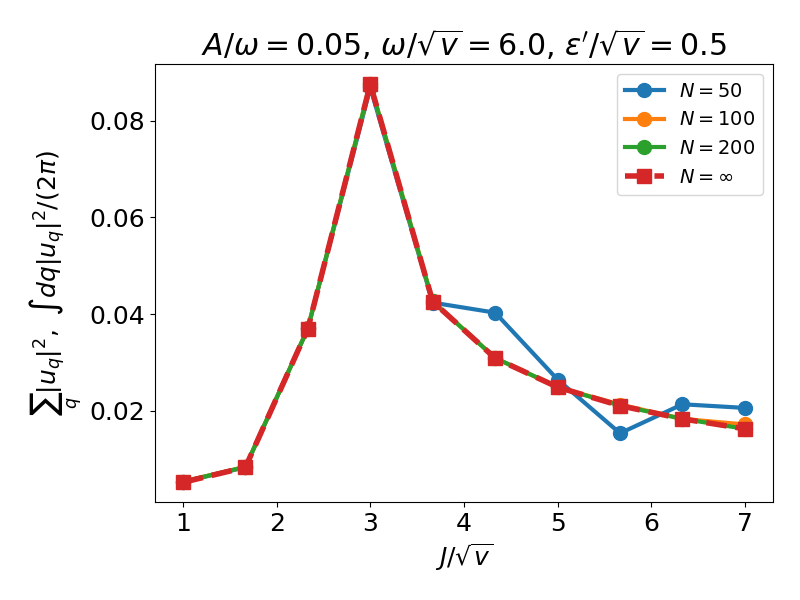}
    \caption{
    Comparison of the integral result for the density of defects due to the non-perturbative effect with the approximate calculation by summing over a finite $N$ for $\eta =0.05,\ \tilde\omega' =6.0,\ \tilde\varepsilon'=0.5$. The solid line corresponds to the sum and the dashed line to the integral numerically. The sum for $N=200$ is consistent with the integral result and is sufficiently large to be considered as the thermodynamic limit.}
    \label{fig:prob_kz_Nint_eta25_tomega25}
\end{figure}

To confirm that the derived equation \eqref{eq:diag_result} is correct as an approximation, we finally check it with numerical calculation when $N$ is finite as shown in the Fig.~\ref{fig:prob_kz_N_eta25_tomega25}. Because the contribution of $n_{\mathrm{FP}}$ is large, it can be seen that the density of defects behaves differently from the case of no oscillations. As discussed, however, the scaling of $\tilde J^{-1}\propto\sqrt{v}$ does not change because this is the same in the non-adiabatic and adiabatic regions. This means that the QKZM is robust to the time-periodic perturbations. 
To verify that the finite $N$ discussed here is sufficiently large, we compared the integral~\eqref{eq:n_fp} with the sum of \eqref{eq:ising_fk} and the result is shown in Fig.~\ref{fig:prob_kz_Nint_eta25_tomega25}. It can be seen that $N=200$ is sufficient to be regarded as the thermodynamic limit.

\subsection{perturbation in the off-diagonal elements}
\label{offdiag}

Next, we consider another time-dependent Hamiltonian
\begin{align}
    H_O(t)&=-\sum^{N}_{j=1}\qty(J(t)\sigma^x_{j}\sigma^x_{j+1}+g(t)\sigma^z_{j})\label{eq:hamiltonian_offdiag},\\
    g(t)&=\frac{1}{4}\qty(vt+\varepsilon'),\\
    J(t)&=\frac{1}{2}\Delta'+\frac{B'}{4}\cos\omega t,
\end{align}
where we impose the periodic boundary condition.

As before, introducing spinless fermions by the JW transformation yields
\begin{align}
    i\frac{d}{dt}\mqty(c_q(t)\\c^\dagger_{-q}(t))&=\mqty( E_q(t)&\delta_q(t)\\\delta_q(t)&- E_q(t))\mqty(c_q(t)\\c^\dagger_{-q}(t)),\\
     E_q(t)&=2J(t)\cos q+\frac{1}{2}\qty(vt+\varepsilon'),\\
    \delta_q(t)&=-2J(t)\sin q.
\end{align}

\begin{figure}[h]
    \centering
    \includegraphics[width=0.95\linewidth]{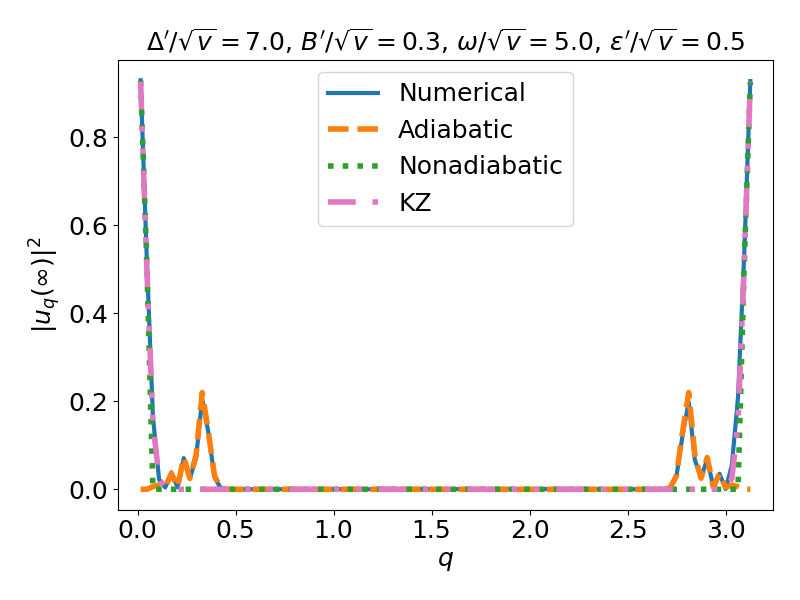}
    \caption{
    Comparison of numerical calculations and approximate expressions of $|u_q(\infty)|^2$ for $\tilde \Delta'=7.0,\ \tilde B' =0.05, \ \tilde\omega=5.0,\ \tilde\varepsilon'=0.5$. It is plotted as $N=200$. The solid line shows the numerical calculation of $|u_q(\infty)|^2$, while the green dotted line is the result of plotting the approximate expression in the non-adiabatic region~\eqref{eq:nonadi_uq_offdiag} and the orange dashed line is the result of plotting the approximate expression in the adiabatic region~\eqref{eq:adi_uq_offdiag}. The pink dash-dot line shows the result without oscillation: $B'=0$. }
    \label{fig:uq_kz_N_Bp10_tomega50}
\end{figure}

First, we consider a non-adiabatic region: $\kappa_q=\tilde \Delta'^2 \sin^2 q\ll1$. In this region, the transition amplitude becomes
\begin{widetext}
\begin{align}
    |u_q(\infty)|^2&\simeq \exp\Biggl(-4\pi\sin^2 q\sum_{n=-\infty}^{\infty}\sum_{m=-\infty}^{\infty}\qty(\tilde\Delta' J_{n}(\eta_B\cos q)+\frac{\tilde B'}{4}J_{n+1}(\eta_B\cos q)+\frac{\tilde B'}{4}J_{n-1}(\eta_B\cos q))\\
    &\qquad \times\qty(\tilde\Delta' J_{m}(\eta_B\cos q)+\frac{\tilde B'}{4}J_{m+1}(\eta_B\cos q)+\frac{\tilde B'}{4}J_{m-1}(\eta_B\cos q)) \\
    &\qquad \times\cos\qty(\tilde\omega\qty(\frac{1}{2}(n^2-m^2)\tilde\omega-(n-m)(\tilde\varepsilon'+2\tilde\Delta'\cos q)))\theta(n-m)\Biggr),\label{eq:nonadi_uq_offdiag}
\end{align}
where we define $\eta_B=B'/\omega$. We note that this expression becomes the same with \eqref{eq:nonadi_diag} in the adiabatic limit if the amplitude is small enough. On the other hand, in the adiabatic region $\kappa_q=\tilde \Delta'^2 \sin^2 q\gg1$, we obtain
\begin{align}
    |u_q(\infty)|^2&\simeq  \pi^2e^{-2\pi\kappa_q}\biggr|\eta_B\cos q {}_1\tilde F_1\qty(-i\kappa_q,0,i\tilde\omega^2)-i\tilde B'\sin q \frac{\sqrt{\kappa_q}}{2}\biggl({}_1\tilde{F}_1\left(-i \kappa_q,1,i\tilde\omega^2\right) +{}_1\tilde{F}_1\left(-i \kappa_q+1,1,i\tilde\omega^2\right)\biggr)\biggr|^2.\label{eq:adi_uq_offdiag}
\end{align}
\end{widetext}
These expressions of $|u_q(\infty)|^2$ have values only at $q\simeq 0,\pm\pi$ if $\tilde\Delta'$ is sufficiently large (Fig.~\ref{fig:uq_kz_N_Bp10_tomega50}), which is easy to see from the asymptotic expansion of $\kappa_q$. Furthermore, the first term in the absolute value in~\eqref{eq:adi_uq_offdiag} is the largest contribution compared to the others because we focus on the regions at $q\simeq 0,\pm\pi$. This shows that the integral of $|u_q(\infty)|^2$ scales with $\tilde \Delta'^{-1}$, as in $H_\mathrm{D}(t)$. From the above discussion, the analytical expression for the density of defects $n(\infty)$ can be expressed as in~\eqref{eq:diag_result}.

\begin{figure}[h]
    \centering
    \includegraphics[width=0.95\linewidth]{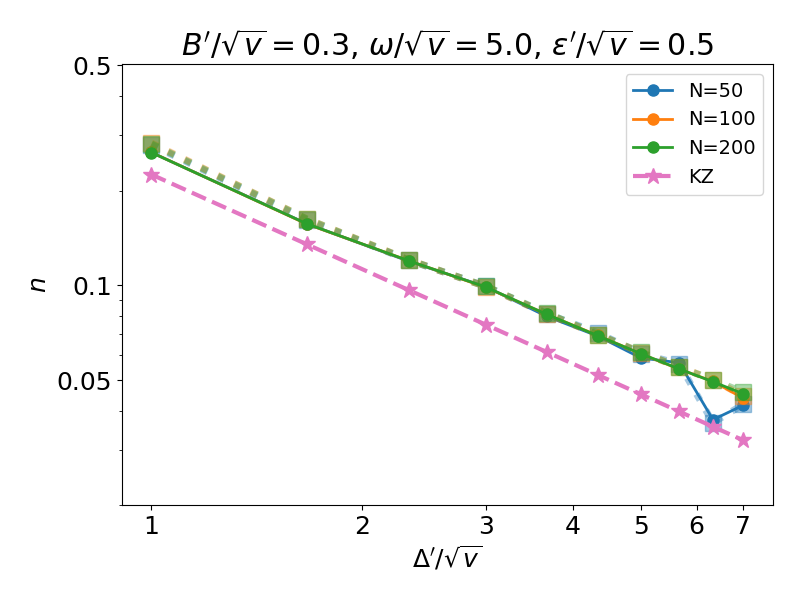}
    \caption{
    Comparison of numerical calculations and approximate expressions of the density of defects in case $N\in\{50,100,200\},\  \tilde B' =0.3,\ \tilde\omega =5.0,\ \tilde\varepsilon'=0.5$. The solid line is the result of solving the Schr\"odinger equation numerically, the dotted line (square) corresponds to the approximate expression~\eqref{eq:diag_result}, and the dashed line (pink) corresponds to the result of the QKZM which is the same in $B=0$. When solving the Schr\"odinger equation numerically, the initial and final times are set to $\tau=-500,500$, respectively. It can be seen that the numerical and approximate results are in good agreement. As in the diagonal oscillation, the density of defects differs from that without oscillation. However, with or without oscillation, both are found to be scaled by $\tilde J^{-1}\propto \sqrt{v}$. This means that the QKZM is robust to the time-periodic perturbation.}
    \label{fig:prob_kz_N_Bp10_tomega10}
\end{figure}

We check it with a numerical calculation when $N$ is finite. 
In the Fig.~\ref{fig:prob_kz_N_Bp10_tomega10}, we compared the density of defects obtained by numerically solving Schr\"odinger equation with an approximate analytical expression. As in the previous subsection, $N=200$ can be regarded as the thermodynamic limit. This figure shows that the density of defects behaves differently compared to the case without oscillations. However, the scaling of the non-perturbative contribution is $\tilde \Delta'^{-1}\propto \sqrt{v}$, which is not different from the scaling predicted by the QKZM, indicating the robustness of the scaling.
In addition, the resonance phenomenon was observed in the diagonal oscillation, but when there was oscillation in the off-diagonal elements, the resonance phenomenon was canceled out by the contribution of the second term in the absolute value of \eqref{eq:adi_uq_offdiag}.

\section{Conclusion\label{sec:conclusion}}

The QKZM is currently attracting attention, and scaling laws for models beyond the simple setting of an isolated system and linear sweep are also of interest. In this paper, we consider a model in which an oscillating external field is perturbatively added in addition to the usual linear linear sweep. In such a setting, it was found that it is necessary to consider not only a perturbative correction term for transitions in the non-adiabatic region, as in the usual QKZM, but also a non-perturbative correction in the adiabatic region. Moreover, although the power spectrum of transition probability is different between with and without oscillation, the non-perturbative correction term also scales as $\sqrt{v}$ in the adiabatic limit, as in the usual QKZM, indicating the robustness of the QKZM with respect to the scaling law.

In the present study, the high symmetry in the model allows for the analytical discussion. The scaling laws of the QKZM have also been investigated for other models such as the spin glass model~\cite{dziarmaga2006dynamics,caneva2007adiabatic,suzuki2011kibble}. The relation between symmetry and the effect of time-periodic perturbations on the QKZM is a subject for future work. Furthermore, we need to investigate the robustness of other quantities, such as kink-kink correlations~\cite{cincio2007entropy,del2018universal,nowak2021quantum,roychowdhury2021dynamics,mayo2021distribution,dziarmaga2022kink}.

\section*{Acknowledgement}
We thank H. Nakazato, M. Fujiwara, and G. Kato for helpful discussions.

\appendix

\section{Furry Picture}\label{appsec:FP}
\begin{widetext}

In this section, we use the FP to obtain the transition probability~\eqref{eq:p_2level_fk}.
In the following discussion, we use these relations
\begin{align}
\int_{-\infty}^\infty d\tau\, e^{i\tilde\omega \tau} D_{\nu_1}\qty(e^{i\frac{\pi}{4}}\tau) D_{\nu_2}\qty(e^{-i\frac{\pi}{4}}\tau)&=\frac{2\pi }{ \Gamma(-\nu_1)}e^{-i\frac{\pi}{4}(\nu_1-\nu_2)}
e^{-i\frac{\tilde\omega^2}{2}}\tilde\omega^{-\nu_1-\nu_2-1}
U\left(-\nu_2,-\nu_1-\nu_2,i\tilde\omega^2\right),\label{eq:formula1}\\
\int_{-\infty}^\infty d\tau\, e^{i\tilde\omega \tau} D_{\nu_1}\qty(e^{i\frac{\pi}{4}}\tau) D_{\nu_2}\qty(e^{i\frac{\pi}{4}}\tau)&=\frac{\sqrt{2\pi} \Gamma(\nu_2+1) }{\Gamma(-\nu_1)}e^{-i\frac{\pi}{4}(\nu_1+3\nu_2+1)}e^{-i\frac{\tilde\omega^2}{2}}\tilde\omega^{-\nu_1+\nu_2}U\left(\nu_2+1,-\nu_1+\nu_2+1,i\tilde\omega^2\right)\\
&\quad +\sqrt{2\pi}\Gamma(\nu_2+1)e^{-i\frac{\pi}{4}(\nu_1-\nu_2+1)}e^{-i\frac{\tilde\omega^2}{2}}\tilde\omega^{-\nu_1+\nu_2}{}_1\tilde{F}_1\left(\nu_2+1,-\nu_1+\nu_2+1,i\tilde\omega^2\right), \label{eq:formula2}\\
\int_{-\infty}^\infty d\tau\, e^{i\tilde\omega \tau} D_{\nu_1}\qty(e^{-i\frac{\pi}{4}}\tau) D_{\nu_2}\qty(e^{-i\frac{\pi}{4}}\tau)&=
\sqrt{2\pi} e^{i\frac{\pi}{4}(\nu_1+3\nu_2+1)}e^{-i\frac{\tilde\omega^2}{2}}\tilde\omega^{-\nu_1+\nu_2}
U\left(-\nu_1,-\nu_1+\nu_2+1,i\tilde\omega^2\right),\label{eq:formula3}
\end{align}
\end{widetext}
where $D_\nu(z)$ is the parabolic cylinder function, ${}_1\tilde F_1(a,b,x)$ is the regularized confluent hypergeometric function of the first kind, and $U(a,b,x)$ is the confluent hypergeometric function of the second kind. We note that these relations are derived from the integral expressions of the special functions~\cite{gradshteyn2014table} and applicable only when $\tilde\omega>0$.

We consider the dimensionless Hamiltonian
\begin{align}
    H(\tau)&= \frac{1}{2}( \tau-\tilde A \cos \tilde \omega \tau+\tilde \varepsilon)\sigma_z+\qty(\tilde \Delta+\frac{\tilde B}{2}\cos\tilde \omega \tau)\sigma_x,
\end{align}
where we define $\tau=\sqrt{v}t$ and $\tilde \circ=\circ/\sqrt{v}$. The time-evolution operator for the LZSM Hamiltoinan ($\tilde A=\tilde B=0$) is
\begin{widetext}
\begin{align}
    U_0(\tau,\tau_0)&=
    \begin{pmatrix}
    f(\tau,\tau_0)&-g^\ast(\tau,\tau_0)\\
    g(\tau,\tau_0)&f^\ast(\tau,\tau_0)
    \end{pmatrix},\\
    f\left(\tau, \tau_0\right) & =e^{-\frac{\pi}{2} \kappa } D_{i \kappa}\qty(e^{-\frac{\pi}{4} i} (\tau_0+\tilde\varepsilon)) D_{-i \kappa}\qty(e^{\frac{\pi}{4} i} (\tau+\tilde\varepsilon)) +e^{-\frac{\pi}{2} \kappa } \kappa D_{-i \kappa-1}\qty(e^{\frac{\pi}{4} i} (\tau_0+\tilde\varepsilon))  D_{i \kappa-1}\qty(e^{-\frac{\pi}{4} i} (\tau+\tilde\varepsilon)) \\
    & =f_1\left(\tau_0\right) D_{-i \kappa}\qty(e^{\frac{\pi}{4} i} (\tau+\tilde\varepsilon))+f_2\left(\tau_0\right) \sqrt{\kappa} D_{i \kappa-1}\qty(e^{-\frac{\pi}{4} i}( \tau+\tilde\varepsilon)),
    \\
    g\left(\tau, \tau_0\right) & =e^{-\frac{\pi}{2} \kappa } e^{\frac{\pi}{4} i}  \sqrt{\kappa} D_{i \kappa}\qty(e^{-\frac{\pi}{4} i} (\tau_0+\tilde\varepsilon)) D_{-i \kappa-1}\qty(e^{\frac{\pi}{4} i} (\tau+\tilde\varepsilon))-e^{-\frac{\pi}{2} \kappa } e^{\frac{\pi}{4} i} \sqrt{\kappa} D_{-i \kappa-1}\qty(e^{\frac{\pi}{4} i}( \tau_0+\tilde\varepsilon))  D_{i \kappa}\qty(e^{-\frac{\pi}{4} i} (\tau+\tilde\varepsilon)) \\
    & =g_1\left(\tau_0\right) \sqrt{\kappa} D_{-i \kappa-1}\qty(e^{\frac{\pi}{4} i} (\tau+\tilde\varepsilon))+g_2\left(\tau_0\right) D_{i \kappa}\left(e^{-\frac{\pi}{4} i}( \tau+\tilde\varepsilon)\right).
\end{align}
We note that 
\begin{align}
    g_1(\tau_0)=e^{i\frac{\pi}{4}}f_1(\tau_0),\quad  g_2(\tau_0)=-e^{i\frac{\pi}{4}}f_2(\tau_0)
\end{align}
hold.

From this, the perturbation term can be written as
\begin{align}
    \hat H_1(\tau)&=U_0^\dagger(\tau,\tau_0) H_1(\tau) U_0(\tau,\tau_0)\\
    &=\frac{1}{2}\cos\tilde\omega \tau\begin{pmatrix}
    f^\ast(\tau,\tau_0)&g^\ast(\tau,\tau_0)\\
    -g(\tau,\tau_0)&f(\tau,\tau_0)
    \end{pmatrix}\qty(-\tilde A\sigma_z+\tilde B\sigma_x)\begin{pmatrix}
    f(\tau,\tau_0)&-g^\ast(\tau,\tau_0)\\
    g(\tau,\tau_0)&f^\ast(\tau,\tau_0)
    \end{pmatrix}\\
    &=-\frac{\tilde A}{2}\cos\tilde\omega \tau
    \begin{pmatrix}
    |f|^2-|g|^2&-2f^\ast g^\ast\\
    -2fg&|g|^2-|f|^2
    \end{pmatrix} +\frac{\tilde B}{2}\cos\tilde\omega \tau
    \begin{pmatrix}
    2\Re (fg^\ast)&-(g^\ast)^2+(f^\ast)^2\\
    f^2-g^2&-2\Re (fg^\ast)
    \end{pmatrix},
\end{align}
where the argument $(\tau,\tau_0)$ was omitted. To obtain the transition probability, we need to calculate
\begin{align}
    \int_{-\infty}^\infty d\tau \hat H_1(\tau).
\end{align}
Hereafter, the argument $(\tau_0)$ is omitted.
When $\tilde B=0$, the diagonal part of the integral becomes
\begin{align}
\int_{-\infty}^\infty d\tau\, (\hat H_1(\tau))_{11} &=-\sqrt{2\pi}\eta\Biggl(\qty(|f_1|^2-|f_2|^2)K_1(\kappa,\tilde\omega)  +\sqrt{\kappa}\Re \Biggl(i f_1 f_2^\ast K_2(\kappa,\tilde\omega)\Biggr)\Biggr),\\
K_1(\kappa,\tilde\omega)&=\sqrt{2\pi}e^{-\frac{\pi\kappa}{2}}\Re\left(\frac{ e^{-i\Omega}}{\Gamma(i\kappa)}U\qty(-i\kappa,0,i\tilde \omega^2)\right),\\
K_2(\kappa,\tilde\omega)&= e^{-\pi \kappa}e^{i\Omega}U\qty(i\kappa,0,-i\tilde\omega^2)+e^{-i\Omega}\Gamma(-i\kappa)\left(\frac{e^{-\pi \kappa}}{\Gamma(i\kappa)}U\left(-i\kappa,0,i\tilde\omega^2\right)-{}_1\tilde F_1\left(-i\kappa,0,i\tilde\omega^2\right)\right),\\
\Omega&=\tilde\omega\tilde\varepsilon+\frac{\tilde\omega^2}{2}.
\end{align}
Similarly, the off-diagonal part of the integral becomes
\begin{align}
\int_{-\infty}^\infty d\tau\, (\hat H_1(\tau))_{21}&=\sqrt{2\pi}\eta e^{i\frac{\pi}{4}}\biggl(-2f_1f_2K_1(\kappa,\tilde\omega)+\frac{i}{2}\sqrt{\kappa}\qty(f_2^2 K_2^\ast(\kappa,\tilde\omega)+ f^2_1 K_2(\kappa,\tilde\omega))\biggr).
\end{align}
In the adiabatic limit $ \left|\left\langle 0\left|U_0(\infty)\right| 0\right\rangle\right|^2=e^{-2\pi\kappa}\simeq 0$, the off-diagonal  part can be written more simply as
\begin{align}
\int_{-\infty}^\infty d\tau\, (\hat H_1(\tau))_{21}&\simeq \sqrt{\frac{\pi\kappa}{2}}\eta e^{-i\frac{\pi}{4}} f_2^2 e^{i\Omega}\Gamma(i\kappa){}_1\tilde F_1^\ast\qty(-i\kappa,0,i\tilde\omega^2).
\end{align}

Next, we consider the case $\tilde A=0$. In this case, the integrals of diagonal part and off-diagonal part becomes
\begin{align}
    \int_{-\infty}^\infty d\tau\, (\hat H_1(\tau))_{11}
    &=\frac{\tilde B}{2}\Re\biggl(e^{-i\Omega}K_3(\kappa,\tilde\omega)+e^{i\Omega}K_6(\kappa,\tilde\omega)\biggr),\\
    \int_{-\infty}^\infty d\tau\, (\hat H_1(\tau))_{21}
    &=\frac{\tilde B}{4}\biggl(e^{-i\Omega}K_4(\kappa,\tilde\omega)+e^{i\Omega}K_7(\kappa,\tilde\omega)-e^{-i\Omega}K_5(\kappa,\tilde\omega)-e^{i\Omega}K_8(\kappa,\tilde\omega)\biggr),
\end{align}
where we define
\begin{align}
    K_3(\kappa,\tilde\omega)&=-i(|f_1|^2-|f_2|^2)\sqrt{\kappa}\frac{2\pi e^{-\frac{\pi}{2}\kappa}}{ \Gamma(i\kappa)}
U\left(-i\kappa+1,1,i\tilde\omega^2\right)\\
    &\quad +\qty(f_1f_2^\ast\frac{ \Gamma(-i\kappa) }{\Gamma(i\kappa)}-f_2f_1^\ast) \kappa 
    \sqrt{2\pi} e^{-\pi\kappa}U\left(-i\kappa+1,1,i\tilde\omega^2\right)  +if_1f_2^\ast\sqrt{2\pi}\Gamma(-i\kappa+1){}_1\tilde{F}_1\left(-i\kappa+1,1,i\tilde\omega^2\right),\\\label{eq:fg_trans}\\
    K_4(\kappa,\tilde\omega)&=\qty(f_1^2\frac{ \Gamma(-i\kappa)}{\Gamma(i\kappa)}-f_2^2)e^{-i\frac{3\pi}{4}}\kappa\sqrt{2\pi}e^{-\pi\kappa}U\left(-i\kappa+1,1,i\tilde\omega^2\right)\\
    &\quad +f_1^2\sqrt{2\pi}\Gamma(-i\kappa+1)e^{-i\frac{\pi}{4}}{}_1\tilde{F}_1\left(-i\kappa+1,1,i\tilde\omega^2\right)+4\pi f_1f_2 \sqrt{\kappa}\frac{ e^{-\frac{\pi}{2}\kappa}e^{-i\frac{\pi}{4}}}{ \Gamma(i\kappa)}
U\left(-i\kappa+1,1,i\tilde\omega^2\right),\label{eq:f^2_trans}\\
    K_5(\kappa,\tilde\omega)&=-f_1^2\sqrt{2\pi}\Gamma(-i \kappa+1)e^{-\frac{i\pi}{4}}{}_1\tilde{F}_1\left(-i \kappa,1,i\tilde\omega^2\right) \\
    &\quad -4i\pi f_1f_2\sqrt{\kappa}\frac{e^{-\frac{\pi}{2}\kappa}e^{-i\frac{\pi}{4}}}{ \Gamma(i \kappa+1)}
U\left(-i \kappa,1,i\tilde\omega^2\right) +\qty(f_2^2+f_1^2 \frac{ \Gamma(-i \kappa) }{\Gamma(i \kappa)})\sqrt{2\pi} e^{-\pi\kappa}e^{i\frac{3\pi}{4}}
U\left(-i\kappa,1,i\tilde\omega^2\right),\label{eq:g^2_trans}\\
    K_6(\kappa,\tilde\omega)&=-i\biggl((|f_1|^2-|f_2|^2)\sqrt{\kappa}\frac{2\pi e^{\frac{-\pi}{2}\kappa}}{ \Gamma(i\kappa+1)}
U\left(-i\kappa,1,i\tilde\omega^2\right) \\
    &\quad +\sqrt{2\pi} e^{-\pi\kappa}\qty(-f_1^\ast f_2+f_2^\ast f_1  \frac{ \Gamma(-i\kappa) }{\Gamma(i\kappa)})U\left(-i\kappa,1,i\tilde\omega^2\right) +f_2^\ast f_1 \sqrt{2\pi}\Gamma(-i\kappa+1){}_1\tilde{F}_1\left(-i\kappa,1,i\tilde\omega^2\right) \biggr)^\ast,
    \label{eq:fg_trans_minus}\\
    K_7(\kappa,\tilde\omega)&=\Biggl(\qty((f_1^2)^\ast+(f_2^2)^\ast\frac{ \Gamma(-i\kappa)}{\Gamma(i \kappa)})
\sqrt{2\pi} e^{-\pi \kappa}e^{i\frac{\pi}{4}}
U\left(-i\kappa,1,i\tilde\omega^2\right)\\
    &\quad +4\pi f^\ast_1f^\ast_2 \sqrt{\kappa}\frac{ e^{-\frac{\pi }{2}\kappa}e^{i\frac{\pi}{4}}}{ \Gamma(i \kappa+1)}
U\left(-i \kappa,1,i\tilde\omega^2\right)+(f_2^2)^\ast\sqrt{2\pi}\kappa\Gamma(-i \kappa)e^{-\frac{i\pi}{4}}{}_1\tilde{F}_1\left(-i \kappa,1,i\tilde\omega^2\right) \Biggr)^\ast,
    \label{eq:f^2_trans_minus}\\
    K_8(\kappa,\tilde\omega)&=\biggl(\kappa\sqrt{2\pi} e^{-\pi \kappa}e^{i\frac{3\pi}{4}}\qty((f_1^2)^\ast+(f_2^2)^\ast\frac{ \Gamma(-i\kappa)}{\Gamma(i\kappa)})
U\left(-i\kappa+1,1,i\tilde\omega^2\right)\\
    &\quad +4i\pi f^\ast_1f^\ast_2\sqrt{\kappa}\frac{ e^{-\frac{i\pi}{4}}e^{-\frac{\pi}{2}\kappa}}{ \Gamma(i\kappa)}
U\left(-i\kappa+1,1,i\tilde\omega^2\right) -i(f_2^2)^\ast\sqrt{2\pi}\Gamma(-i\kappa+1)e^{-\frac{i\pi}{4}}{}_1\tilde{F}_1\left(-i\kappa+1,1,i\tilde\omega^2\right) \biggr)^\ast.\label{eq:g^2_trans_minus}
\end{align}
In the adiabtic limit, the off-diagonal part becomes
\begin{align}
    \int_{-\infty}^\infty d\tau(\hat H_1(t))_{21}
    &\simeq \frac{\tilde B}{4}f_2^2e^{i\Omega}\sqrt{2\pi}\kappa\Gamma(i \kappa)e^{i\frac{\pi}{4}}\biggl({}_1\tilde{F}_1\left(-i \kappa,1,i\tilde\omega^2\right)  + {}_1\tilde{F}_1\left(-i\kappa+1,1,i\tilde\omega^2\right)  \biggr)^\ast.
\end{align}
\end{widetext}

From the above discussion, the transition probability in the adiabatic limit under $A\neq 0, B\neq 0$, and $t_0\to-\infty$ becomes
\begin{align}
    & |\langle \uparrow|U(\infty)| \uparrow\rangle|^2 \\
    &\simeq \left|\bra{\uparrow}U_0(\infty)\ket{\uparrow}-i\bra{\uparrow}U_0(\infty)\int_{-\infty}^{\infty} d t \tilde{H}_1(t)\ket{\uparrow}\right|^2 \label{eq:prob_fk_exact}\\
    &\simeq \left|\bra{\downarrow}\int_{-\infty}^{\infty} d t \tilde{H}_1(t)\ket{\uparrow}\right|^2 \\
    &\simeq \frac{\pi\kappa}{2}e^{-\pi\kappa}|\Gamma(i\kappa)|^2\biggr|\eta {}_1\tilde F_1^\ast\qty(-i\kappa,0,i\tilde\omega^2)\\
    &\quad +i\tilde B \sqrt{\kappa}\frac{1}{2}\biggl({}_1\tilde{F}_1\left(-i \kappa,1,i\tilde\omega^2\right)  + {}_1\tilde{F}_1\left(-i\kappa+1,1,i\tilde\omega^2\right)  \biggr)^\ast\biggl|^2\\
    &\simeq \pi^2e^{-2\pi\kappa}\biggr|\eta {}_1\tilde F_1\qty(-i\kappa,0,i\tilde\omega^2)\\
    &\quad -i\tilde B \frac{\sqrt{\kappa}}{2}\biggl({}_1\tilde{F}_1\left(-i \kappa,1,i\tilde\omega^2\right)  + {}_1\tilde{F}_1\left(-i\kappa+1,1,i\tilde\omega^2\right)  \biggr)\biggl|^2,
\end{align}
where we use these relation
\begin{align}
    f_1(\tau_0)&\to
    e^{-\frac{5\pi}{4}\kappa}e^{\frac{i}{4}|\tau_0|^2}|\tau_0|^{i\kappa},\\
    f_2(\tau_0)&\to
    e^{-\frac{\pi}{4}\kappa} \sqrt{1-e^{-2\pi\kappa}}e^{i\arg\Gamma(1-i\kappa)} 
    e^{\frac{i}{4}|\tau_0|^{2}}|\tau_0|^{i\kappa},
\end{align}
which hold in $\tau_0\to-\infty$.

\bibliography{ref} 

\end{document}